\documentclass[fleqn,10pt]{wlscirep}
\usepackage[utf8]{inputenc}
\usepackage[T1]{fontenc}
\usepackage{amsmath, braket} 
\usepackage{amssymb} 
\usepackage{graphicx} 
\usepackage{ulem, bbold}
\usepackage{hyperref} 
\usepackage{comment}
\usepackage{tikz}  
\usepackage{caption}
\usepackage{subcaption}
\usetikzlibrary{arrows,shapes,positioning,shadows,svg.path,backgrounds,fit}

\newcommand{\op}[1]{\hat{#1}}

\DeclareMathOperator\atanh{arctanh}

\title{Measurement Induced Synthesis of Coherent Quantum Batteries}

\author[1]{Mariia Gumberidze}
\author[1,*]{Michal Kol\'a\v r}
\author[1]{Radim Filip}
\affil[1]{Department of Optics, Palack\'y University, 17. listopadu 12, 771 46 Olomouc, Czech Republic}

\affil[*]{kolar@optics.upol.cz}

\begin{abstract}
Quantum coherence represented by a superposition of energy eigenstates is, together with energy, an important resource for quantum technology and thermodynamics. Energy and quantum coherence however, can be complementary. The increase of energy can reduce quantum coherence and vice versa. Recently, it was realized that steady-state quantum coherence could be autonomously harnessed from a cold environment. We propose a conditional synthesis of $N$ independent two-level systems (TLS) with partial quantum coherence obtained from an environment to one coherent system using a measurement able to increase both energy and coherence simultaneously. The measurement process acts here as a Maxwell demon synthesizing the coherent energy of individual TLS to one large coherent quantum battery. The measurement process described by POVM elements is diagonal in energy representation and, therefore, it does not project on states with quantum coherence at all. We discuss various strategies and their efficiency to reach large coherent energy of the battery. After numerical optimization and proof-of-principle tests, it opens way to feasible repeat-until-success synthesis of coherent quantum batteries from steady-state autonomous coherence.       
\end{abstract}
\begin{document}

\flushbottom
\maketitle

\thispagestyle{empty}

\section*{Introduction}

Quantum thermodynamics is an emerging field \cite{millen, kammerlander}, which combines the principles of two different and very successful fields of physics. Namely, quantum mechanics and thermodynamics. From the side of quantum mechanics, it involves quantum superposition and, therefore, quantum coherence. On the other hand, from thermodynamics, it keeps the main interest in energy and methods on how to transform it. Therefore, a preferred representation in quantum thermodynamics is the energy basis. The latter makes it conceptually different from quantum information, which is basis independent. Already in the simplest two-level system (TLS), energy difference and quantum coherence are represented by non-commuting operators. Their values cannot be simultaneously arbitrarily high for a single system, and also their uncertainties are limited by fundamental commutation relations. Recent research interest is focused on various devices, operating on a quantum scale, e.g., quantum batteries \cite{alicki2013,binderNJP2015,binderPRE2015,campaioliPRL2017}, quantum heat engines \cite{ElouardPRL2018} or refrigerators \cite{millen, QMC}, and energy extractability from these devices \cite{llobet2015}. 

Our main focus is on systems with coherent energy, i.e., systems which can appear in a superposition of energy eigenstates being the basic resource of modern quantum physics and, therefore, the viable field of research\cite{LostaglioPRX2015,KwonPRL2018,HenaoPRE2018}.
Such superposition in quantum systems, like atoms and solid-state objects, can be induced by strong coherent force represented, for example, by electric and magnetic fields or maser/laser radiation. On the other hand, it has been recently proposed that a residual steady-state coherent energy (energy of the superposition of energy eigenstates) can be autonomously harnessed by elementary TLS from a sufficiently cold environment \cite{giacomoPRL2018}. Stimulated by these results, we consider such an environmentally-induced coherence as a resource to be synthesized. Therefore, we aim here for synthesizing a certain number $N$ of uncorrelated two-level systems into the higher-dimensional coherent quantum battery. The battery is usually defined as a physical system able to harness the energy from its surroundings\cite{binderNJP2015}, to store it in some appropriate form and, eventually, to release the energy to the surroundings again, usually referred to as the {\it charging, storing} and {\it discharging} processes, respectively. Recent literature describes the manipulation of quantum systems, frequently non-interacting TLSs\cite{FerraroPRL2018}, or interacting chains of TLSs\cite{LePRA2018}, to investigate the effects of the quantum properties (namely entanglement and correlations) of the battery states on the speed of charging and the energy extractability. While the majority of works aim for incoherent charging and discharging (regardless of the resulting coherence of the final state)\cite{binderNJP2015,campaioliPRL2017,LePRA2018}, some recent papers\cite{ManzanoPRE} follow the goal of coherent charging of the battery, meaning to create the final battery state in a superposition of battery energy eigenstates. All these approaches do use some CPTP (completely positive trace-preserving) transformations for the charging protocol, thus are inherently deterministic. 

In the context of quantum thermodynamics there are many positive effects predicted, using coherence as a resource, especially the coherence between non-degenerate energy eigenstates. These could be extracting larger average amount of energy from a quantum system\cite{SkrzypczykNature2014,kammerlander}, increasing the potential usefulness of the thermal reservoirs\cite{HardalSciRep2015,ManzanoPRE2016} or increase in the power output of thermal machines\cite{BrunnerPRE2014,UzdinPRX2015}, supported by an experimental demonstration\cite{KlatzowPRL2019}. 

To synthesize a higher dimensional quantum coherent battery from the independent TLSs, Maxwell's demon principle can be advantageously used. It is a conditional procedure based on the measurement of a system\cite{greiner}. For increasing the energy only, it would be sufficient to measure TLSs individually in the energy representation and keep only excited systems in a battery, whereas the de-excited systems will be brought once more into interaction with a thermal bath and reused until all of them have maximal energy (the so-called repeat-until-success strategy). Efficient energy measurement is, therefore, a thermodynamic resource\cite{ToyabeNature2010,KoskiPNAS13786,KoskiPRL2014,WalmsleyPRL2016,CottetPNAS7561,ElouardPRL2018,QMC,LutzPRL2009}. However, for keeping coherence, the energy measurement is counterproductive since it causes loss of all system coherence. Therefore, to synthesize two-level systems with a partial coherence into a large coherent system requires a different measurement strategy. However, we still demand it to use only measurement elements {\it diagonal} in the TLS energy basis. This fact ensures that the measurement does not directly project on a superposition of energy eigenstates and that it is not a source of energy on its own~\cite{ArakiPhysRev1960,PopescuPRL2014}.          

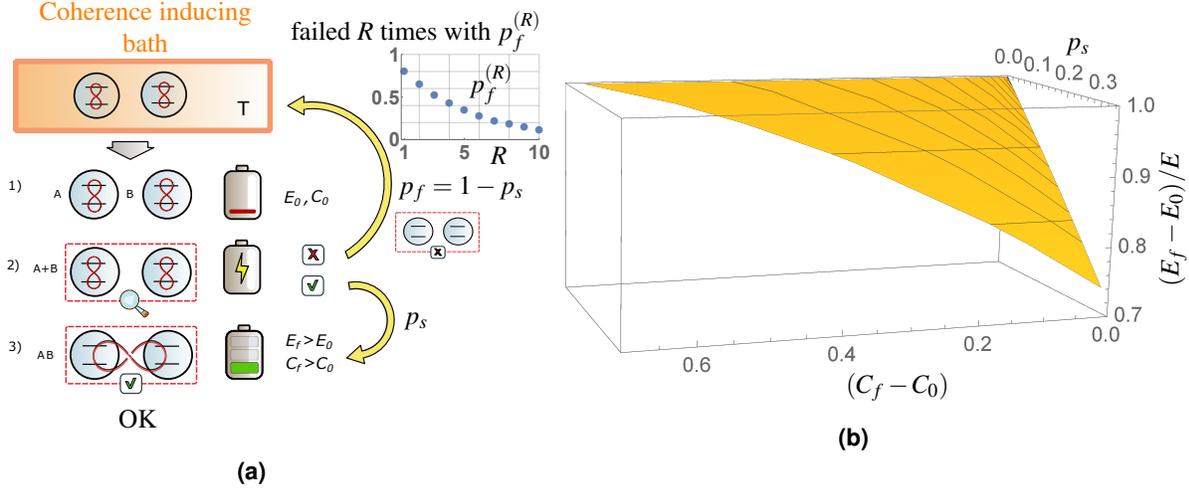
\begin{figure}[ht]
\begin{subfigure}{0.45\textwidth}
\begin{tikzpicture} 
  \node (img1)  {\includegraphics[width=.9\columnwidth]{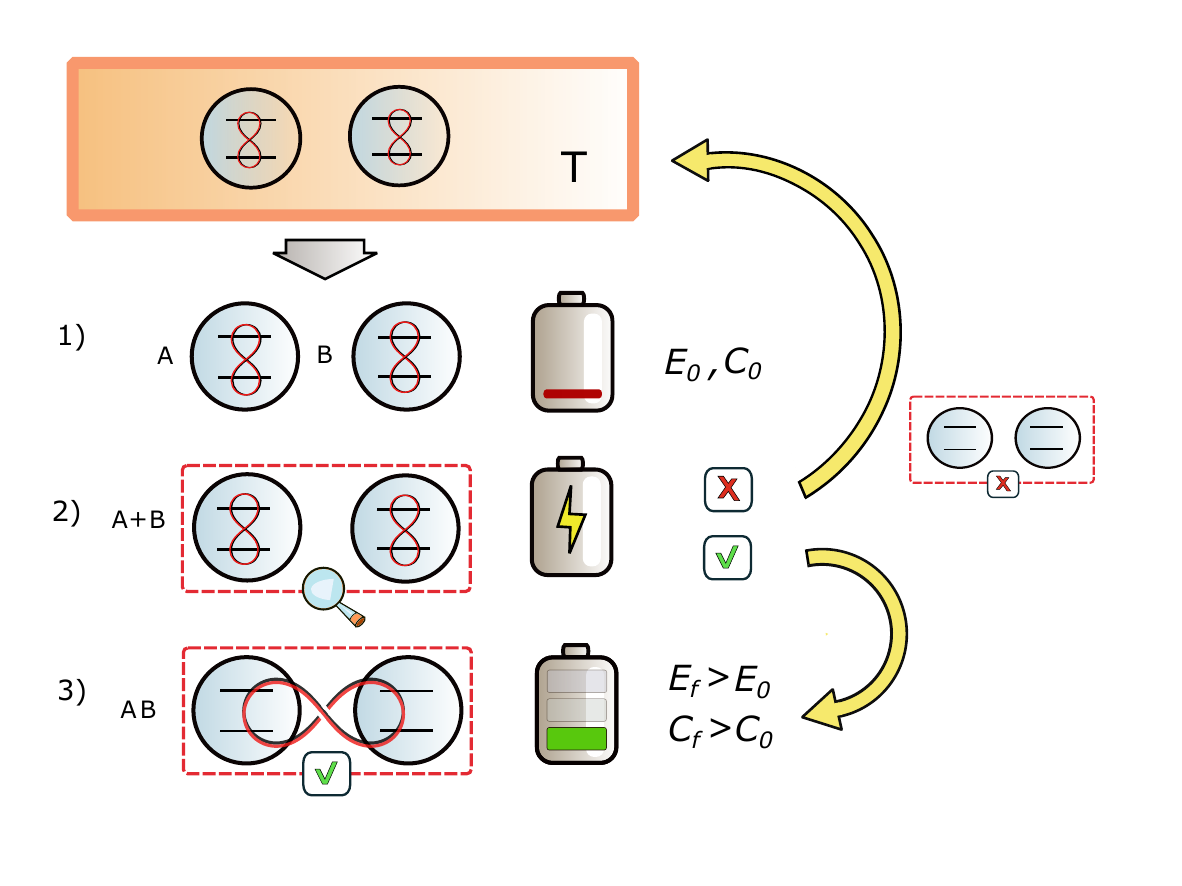}};
  \node[above=of img1, node distance=0cm, yshift=-3.0cm,xshift=2.8cm] (img2)  {\includegraphics[width=.3\linewidth]{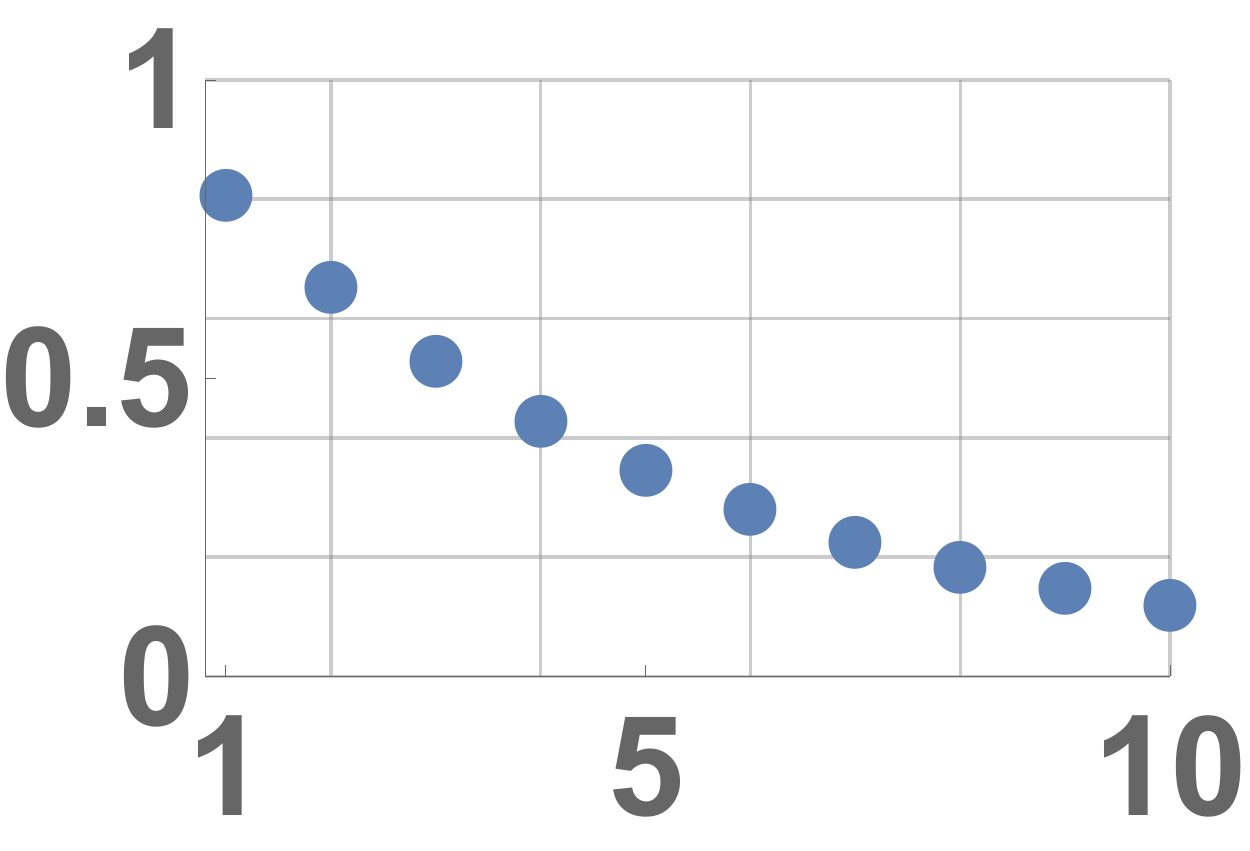}};
   \node[above=of img1, node distance=0cm, yshift=-2.2cm,xshift=3.2cm] {$p_f^{(R)}$};
    \node[above=of img1, node distance=0cm, yshift=-3.0cm,xshift=3.3cm] {$R$};
  \node[above=of img1, node distance=0cm, yshift=-6.5cm,xshift=-1.5cm] {OK};
   \node[above=of img1, node distance=0cm, yshift=-5.2cm,xshift=2.2cm] {$p_s$};
 \node[above=of img1, node distance=0cm, yshift=-3.5cm,xshift=2.8cm] {$p_f=1-p_s$};
 \node[above=of img1, node distance=0cm, yshift=-1.5cm,xshift=2.2cm] {failed $R$ times with $p_f^{(R)}$};
 \node[above=of img1, node distance=0cm, yshift=-1.5cm,xshift=-1.4cm] {\begin{minipage}{.4\textwidth}
 \begin{center}
 {\color{orange}Coherence inducing bath}
 \end{center}
 \end{minipage}};
\end{tikzpicture}
\centering \hspace{-0.06\linewidth}
\caption{\label{fig-scheme-idea}}
\end{subfigure}
\begin{subfigure}{0.45\textwidth}
\begin{tikzpicture} 
  \node (img1)  {\includegraphics[width=.99\columnwidth]{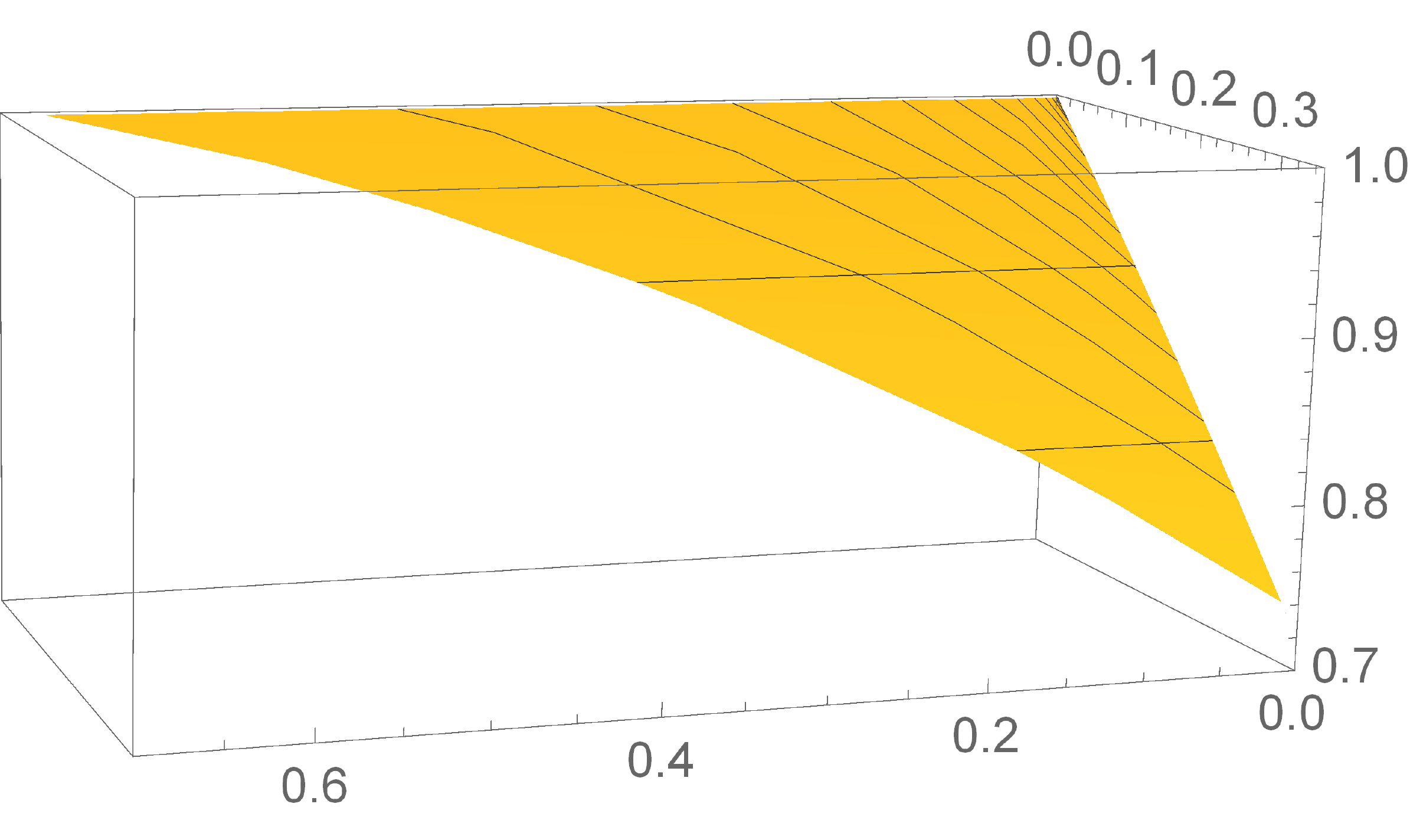}};
  \node[above=of img1, node distance=0cm, yshift=-1.5cm,xshift=2.9cm] {$p_s$};
   \node[above=of img1, node distance=0cm, yshift=-6.1 cm,xshift=.5cm] { $(C_f-C_0)$};
  \node[left=of img1, node distance=0cm, rotate=90, anchor=center,yshift=-9.2cm,xshift=-0.1cm] {$(E_f-E_0)/E$};
\end{tikzpicture}
\caption{\label{fig-param-plot}}
\end{subfigure}
\caption{ (\subref{fig-scheme-idea}) Schematics showing the idea of the 2-TLS (two-level systems) battery charging process in three steps and (inset) the total probability of failure $p_f^{(R)}$ after $R$ unsuccessful repetitions. In the initial step 1) the pair of TLS leaves the cold coherence inducing bath\cite{giacomoPRL2018} and enters the process, assumed to be almost discharged with low initial energy $E_0$ and coherence $C_0$ (red loops connecting the levels). Such bath enables, due to the structure\cite{giacomoPRL2018} of the system-bath interaction
Hamiltonian, the formation of coherence in energy basis of a generic two-level system, independently of its initial state. Step 2) represents the {\it conditional} stage of the charging operation. If the outcome of this step is successful (see step 3) with the single run probability of success $p_s$, the battery is made more coherent and simultaneously charged, i.e., has higher energy $E_f>E_0$ and increased coherence $C_f>C_0$ as well, both with respect to the eigenbasis of Hamiltonian \eqref{eq-Hamiltonian}. If the charging fails, with the single run probability $p_f=(1-p_s)$ in step 2), the battery is completely discharged and incoherent. In principle it can be recycled by bringing it in contact with the cold coherence-inducing bath\cite{giacomoPRL2018} again and the protocol is repeated until success (RUS). If the charging does not successfully occur in $R$ recycling runs, the total probability of failure, $p_f^{(R)}=(1-p_s)^R$, is exponentially converging to zero with the number of repetitions, see the inset for a typical behavior. (\subref{fig-param-plot}) Parametric 3D plot of the relative entropy of coherence difference $\Delta C$, Eq.~\eqref{eq-DeltaC}, on $x$-axis, the normalized average energy difference $\Delta E/E$, Eq.~\eqref{eq-DeltaE}, on $z$-axis and single run probability of success $p_s$ on $y$-axis. 
It demonstrates the basic step in the presented protocol (a): increase in the energy and coherence of the battery (after the charging process) with sufficient probability of success.} 
\end{figure}

In this paper, we propose such strategy which is coherently synthesizing excited two-level systems to a larger coherent system with the possibility to enhance both energy and coherence on two and more TLS by the {\it collective} quantum measurement with elements diagonal in the energy basis. Such synthesizing unifies the battery build-up and charging process. 
As we find out the outcome of the measurement, we acquire information about the battery state and, based on it, we decide on the next step of the protocol, see Fig.~\ref{fig-scheme-idea}. If the TLSs did not synthesize as required, we let them interact with the thermal bath surrounding again to provide once more residual coherent energy, and repeat it until the success. In this way, we can asymptotically approach deterministic charging at a cost of the number of TLS interacting with the bath. 

Our main result shows the possibility of elementary coherent energy synthesizing, i.e., the increase of both the initial energy and coherence \cite{baumgratz} (both small but nonzero) employing quantum measurement on the system. This result is quantitatively presented in Fig.~\ref{fig-param-plot}. Such energy and coherence increasing measurement can be extended and optimized further, as we subsequently propose, to reach maximal coherent energy of the battery. 
Although this charging stage is, in a single protocol run, conditional with a limited probability of success $p_s$, it can be repeated until it succeeds and TLS are synthesized and coherently charged with the overall success probability arbitrarily close to one. Our work represents a bottom-to-top approach and analyzes only the basic recycling strategy. 
We generalize the protocol to higher number $N$ of TLS, although sticking with the general idea of global measurement on these copies, and open the road for future numerical optimization in this direction, as well. Top-to-bottom numerical approach based purely on numerical optimization requires the insight presented here anyway. 


As our focus is on coherent energy synthesis, we consider all other incoherent energy operations, e.g., energy increase without coherence change, to be without any cost. The information, obtained in our measurement-based protocol, is physical, thus the measurement result has to be stored in some classical physical register (memory) which, after its filling, has to be refreshed with nonzero incoherent energy cost, dictated by Landauer's principle\cite{LutzPRL2009}. We do not explicitly consider such necessary incoherent energy resources in this paper, as at this stage we are primarily interested in the effect of coherent energy synthesis, and not so much in classical and incoherent resources allowing its functionality.

\section*{Coherently charging a pair of two-level systems} 
\label{sec-2-systems}

 In the following subsection, we describe the system and the variables of interest. The next subsections introduce the main idea of the paper, i.e., the conditional charging procedure applied to a pair of TLS in pure initial states, showing also which projective measurement is optimal concerning our goal. The following subsection generalizes the results for two classes of partially coherent states, namely, for partially mixed states due to dephasing and spontaneous emission of energy from TLS.

\subsection*{Energy and Coherence}
\label{sub-systems}
As the system, we consider two independent copies of two-level systems (TLS), denoted $1$ and $2$, with a non-zero energy gap $E$ between the states. The preferred basis of the pair of TLSs, which we will uniquely refer to throughout the paper, will be given by the TLS energy eigenstates labeling the ground and excited state, respectively, $\{\ket{g_j},\ket{e_j}\}$, for both subsystems $j=1,2$.  
The Hamiltonians defining these states read 
\begin{eqnarray}
\op{H}_j=\frac{E}{2}\left(\ket{e_j}\bra{e_j}-\ket{g_j}\bra{g_j}\right),\quad j=1,2,
\label{eq-Hams}
 \end{eqnarray}
where $E$ is the energy gap of each TLS, yielding the Hamiltonian of the total system
\begin{eqnarray}
\op{H}=\hat{H}_{1}\otimes\hat{1}_{2}+\hat{1}_{1}\otimes\hat{H}_{2}.
\label{eq-Hamiltonian}
\end{eqnarray}
From an energetic point of view, the Hamiltonian \eqref{eq-Hamiltonian} represents a four-level system with the following energy eigenstates: $\ket{g_1g_2}$ ground state with energy $-E$, doubly degenerate first excited states $\{\ket{e_1g_2},\ket{g_1e_2}\}$ with zero energy, and  $\ket{e_1e_2}$ with energy $+E$. All the energy considerations will be made relative to this total Hamiltonian~\eqref{eq-Hamiltonian}. The average energy and the variance of energy of any quantum state of our system is determined in a standard way as
\begin{equation}
\langle E\rangle={\rm Tr}(\op{\rho}\op{H}), \quad \langle \Delta E^2\rangle={\rm Tr}(\op{\rho}\op{H}^2)-\langle E\rangle^2.
\label{eq-E}
\end{equation}

To quantify coherence of the state, we use the relative entropy of coherence  (denoted simply "coherence" from now on) defined as  \cite{baumgratz} 
\begin{equation}
C(\hat{\rho})= S(\hat{\rho}_{diag})- S(\hat{\rho}),
\label{eq-rel-ent-coh}
\end{equation}
where $S(\hat{\rho})=-{\rm Tr}(\op{\rho}\ln\op{\rho})$ is von Neumann entropy and $S(\hat{\rho}_{diag})$ is the entropy of the {\it diagonal} version of the state with respect to the {\it energy} basis. Thus, the latter one effectively amounts to the Shannon entropy of state $\hat{\rho}$ concerning energy eigenbasis. The maximum value of coherence $C$ depends on the dimension $D$ of the system Hilbert space. For any system $0\leq C\leq \ln{D}$ holds\cite{baumgratz}, thus, e.g., for a single TLS it is upper bounded by $C=\ln{2}$. If the single basic step of the protocol increases both coherence and energy, this result would constitute a prerequisite to use such coherent energy for an iterative procedure on many TLSs.  

The choice of the relative entropy of coherence, Eq.~\eqref{eq-rel-ent-coh}, as our measure is definitely not unique. As another possibility, we may use as well the $l_1$-norm based measure\cite{baumgratz}. Although both are proper coherence measures, we prefer relative entropy of coherence. Such preference is motivated by the recognized connection of $C(\op{\rho})$ with thermodynamic quantities\cite{kammerlander} and the notion of von Neumann entropy, naturally connecting this measure and thermodynamics in much broader sense. 

As we are interested in changes of the battery energy and coherence, we focus on the behavior of the energy difference 
\begin{equation}
\Delta E=E_f-E_0,\quad E_f\equiv {\rm Tr}(\op{\rho}_f\op{H}),\quad E_0\equiv {\rm Tr}(\op{\rho}_0\op{H}),
\label{eq-DeltaE}
\end{equation}
between the final energy after the charging and the initial energy  before the charging process, respectively. Similarly, we focus on the change of the coherence 
\begin{equation}
\Delta C=C_f-C_0,\quad C_f\equiv C(\op{\rho}_f),\quad C_0\equiv C(\op{\rho}_0),
\label{eq-DeltaC}
\end{equation}
of the battery state.

By advancing substantially presentation of the results, we can see the opening result of this paper plotted in Fig.~\ref{fig-param-plot}. The plot shows the simultaneous increase in the energy $\Delta E>0$ and coherence $\Delta C>0$ of the battery after charging process with non-negligible probability of success $p_s$.

\subsection*{Synthesizing the 2-TLS coherent battery made from pure states }
\label{sub-2-pure}
Before we analyze more realistic cases of TLS in mixed states, we discuss the ideal case showing the main idea of our approach. We consider charging of the battery consisting of a pair of TLS  in the following initial state
\begin{equation}
\ket{\psi_j}=\sqrt{p}\;\ket{e_j}+\sqrt{1-p}\;\ket{g_j},\quad j=1,2,
\label{eq-psi-pure}
\end{equation}
where the Hamiltonians defining the respective energy eigenstates are in Eq.~\eqref{eq-Hams}. This type of state can be, in principle, a steady state of TLS, originating from a certain type of interaction with a thermal bath\cite{giacomoPRL2018}. This bath can increase the TLS energy {\it and} coherence with respect to its ground state $\ket{g}$.
Such case will serve below as an optimization target for harnessing of coherent energy from the environment. The corresponding {\it initial} state of the TLS pair is 
\begin{eqnarray}\label{eq-Psi-init}
\ket{\Psi_i}=\ket{\psi_1}\otimes\ket{\psi_2}=p\ket{e_1e_2}+\sqrt{p(1-p)}\left(\ket{e_1g_2}+\ket{g_1e_2}\right)+(1-p)\ket{g_1g_2}.
\end{eqnarray}

As it is well known, projective measurement applied to the system can change the state closer to a desired one. We want to choose such measurement that will charge the battery, i.e., increase its energy but also coherence, if possible. Moreover, we want to achieve this goal without using projectors on states with any coherence with respect to the energy eigenbasis of the Hamiltonian \eqref{eq-Hamiltonian}. 
If we project on the superposition of energy eigenstates, coherence can be induced by the measurement itself. Therefore, only projectors diagonal in the measurement basis are allowed. 
Namely, we choose the pair $\{\op{P}_0,\op{P}_1=\op{1}-\op{P}_0 \}$, with
\begin{equation}
\op{P}_0=\ket{g_1g_2}\bra{g_1g_2},\quad \op{P}_1=\op{1}-\op{P}_0.
\label{eq-2-projector}
\end{equation}
Application of such measurement results in two conditional post-measurement states
\begin{eqnarray}
\ket{\Psi_0}&=&\frac{\op{P}_0\ket{\Psi_i}}{\sqrt{1-p_s}}=\ket{g_1g_2}, \label{eq-pure-nonsucc}\\
\ket{\Psi_f}&=&\frac{\op{P}_1\ket{\Psi_i}}{\sqrt{p_s}}=\frac{p\ket{e_1e_2}+\sqrt{p(1-p)}\left(\ket{e_1g_2}+\ket{g_1e_2}\right)}{\sqrt{p(2-p)}},\quad p\neq 0, \label{eq-pure-succ}
\end{eqnarray} 
where $p_s=p(2-p)$ is the probability of success of the charging procedure, applied to a pair of TLS, serving as a new normalization factor. Equation~\eqref{eq-pure-succ} represents the {\it final} state after {\it successful} round of the battery charging process. In the case of the success, we fuse the initially independent pair of TLS into a larger superposition, thus creating and coherently charging the battery simultaneously. The failure of the protocol results in reducing the initial energy and coherence to zero, by bringing the TLS pair to the energetic ground state. Thus, these pairs (failed to be charged) can be used once more to harness a partial coherence from the environment. If such a strategy is repeated, the probability of failure after $R$ independent repetitions (meaning that the battery will be not charged even in a single event) decreases as 
\begin{eqnarray}
p_f^{(R)}=(1-p_s)^{R}=(1-p)^{2R},\quad R\geq 1,
\label{eq-pure-RUS-pf}
\end{eqnarray}
with $p$ the probability of single TLS excitation, Eq.~\eqref{eq-psi-pure}. As illustrated in the inset of Fig.~\ref{fig-scheme-idea}, such event becomes almost impossible at an exponential rate.

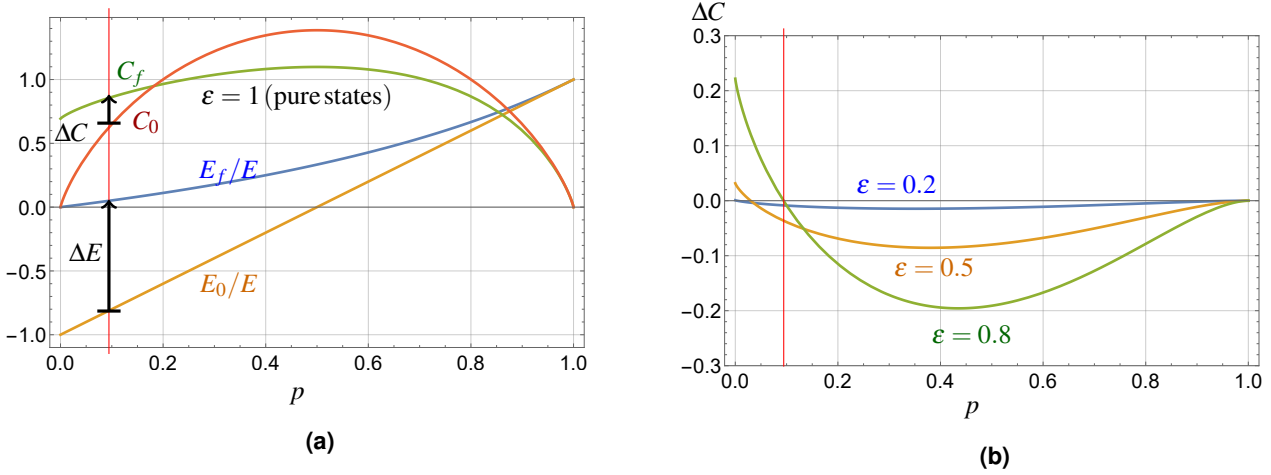
\begin{figure}[ht]
\begin{subfigure}{0.49\textwidth}
\begin{tikzpicture} 
  \node (img1)  {\includegraphics[width=.9\columnwidth]{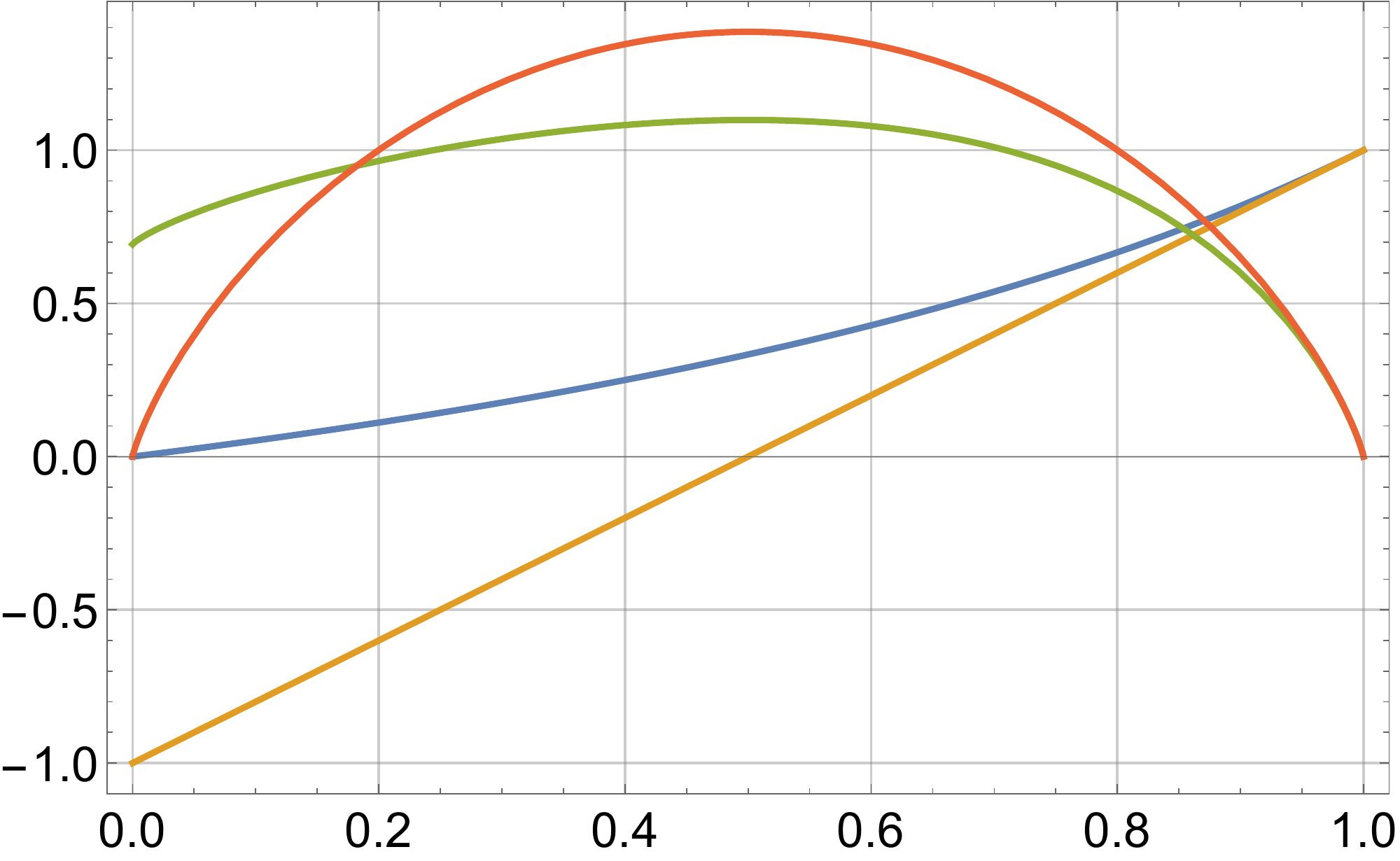}};
  \draw[red,thin] (-2.5,-2.1) -- (-2.5,2.5);
  \draw[black,very thick,|->] (-2.5,.95) -- (-2.5,1.35);
  \draw[black,very thick,|->] (-2.5,-1.55) -- (-2.5,-.05);
  \node[above=of img1, node distance=0cm, yshift=-6.4cm,xshift=0cm] {$p$};
  \node[above=of img1, node distance=0cm, yshift=-2.5cm,xshift=-0.0cm] {$\epsilon =1\,{\rm(pure\, states)}$};
   \node[above=of img1, node distance=0cm, yshift=-2.8cm,xshift=-2.cm] {{\color{black!40!red}$C_0$}};
   \node[above=of img1, node distance=0cm, yshift=-2.9cm,xshift=-3.0cm] {{\color{black}$\Delta{C}$}};
   \node[above=of img1, node distance=0cm, yshift=-4.5cm,xshift=-2.8cm] {{\color{black}$\Delta{E}$}};
   \node[above=of img1, node distance=0cm, yshift=-2.18cm,xshift=-2.2cm] {{\color{black!60!green}$C_f$}};
    \node[above=of img1, node distance=0cm, yshift=-3.5cm,xshift=-0.9cm] {{\color{blue}$E_f/E$}};
    \node[above=of img1, node distance=0cm, yshift=-5.cm,xshift=-0.9cm] {{\color{black!20!orange}$E_0/E$}};
\end{tikzpicture}
\caption{}
\label{fig-2-qub}
\end{subfigure}
\hfill
\begin{subfigure}{0.49\textwidth}
\begin{tikzpicture} 
  \node (img1)  {\includegraphics[width=.9\columnwidth]{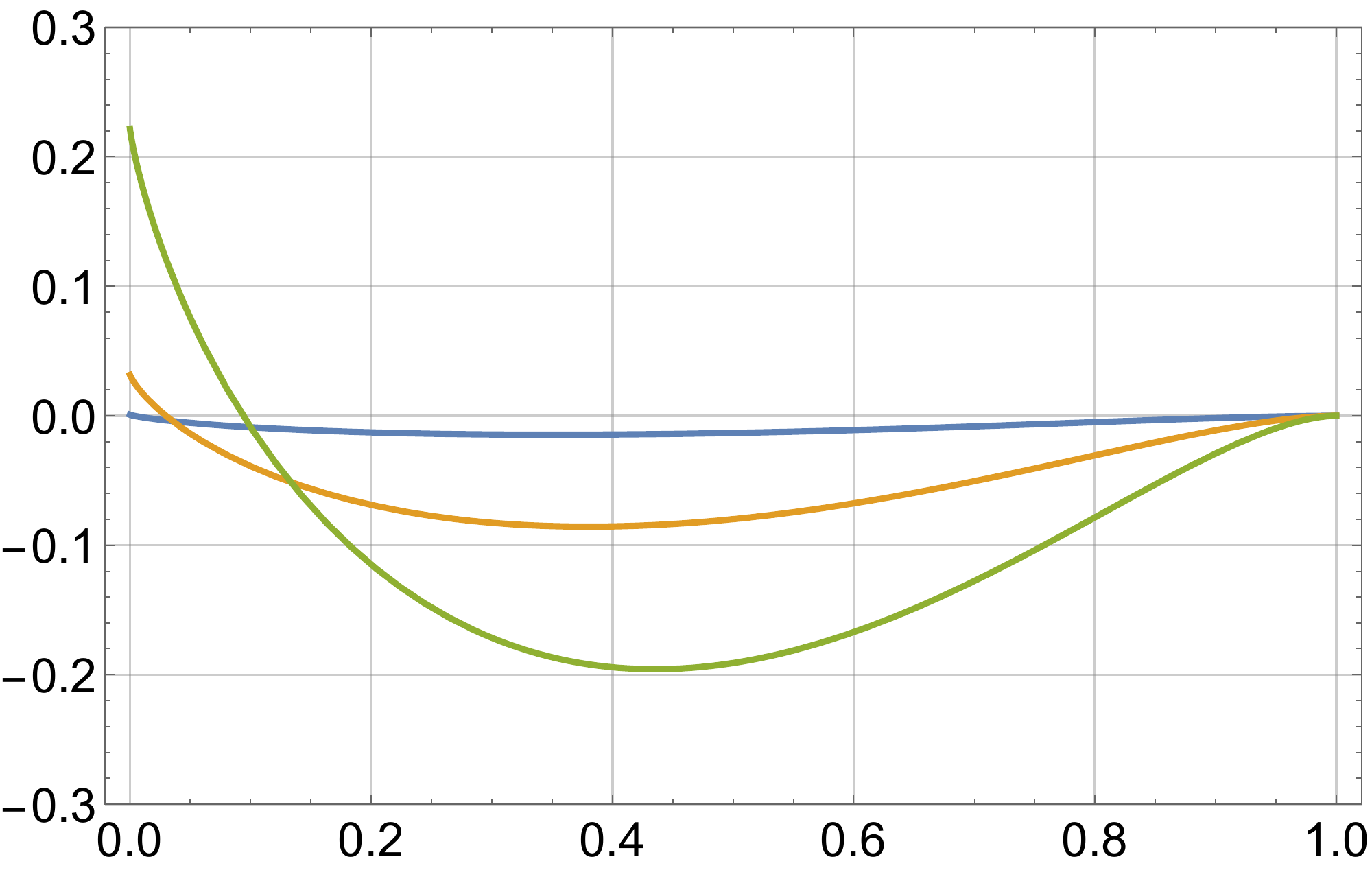}};
  \draw[red,thin] (-2.5,-2.1) -- (-2.5,2.4);
  \node[above=of img1, node distance=0cm, yshift=-6.5cm,xshift=0cm] {$p$};
   \node[above=of img1, node distance=0cm, yshift=-1.2cm,xshift=-3.5cm] {$\Delta C$};
   \node[above=of img1, node distance=0cm, yshift=-3.5cm,xshift=-1.cm] {{\color{blue}$\epsilon=0.2$}};
    \node[above=of img1, node distance=0cm, yshift=-4.6cm,xshift=-0.5cm] {{\color{black!20!orange}$\epsilon=0.5$}};
    \node[above=of img1, node distance=0cm, yshift=-5.5cm,xshift=-0.cm] {{\color{black!60!green}$\epsilon=0.8$}};
\end{tikzpicture}
\caption{}
\label{fig-2-qub-eps}
\end{subfigure}
\caption{ (\subref{fig-2-qub}) The plot of the normalized average energy $E_f/E$, $E_0/E$ and the relative entropy of coherence $C_f$, $C_0$ for the final state $\ket{\Psi_f}$, Eq.~\eqref{eq-pure-succ}, and the initial state $\ket{\Psi_i}$, Eq.~\eqref{eq-Psi-init}, for a pair of TLS in pure states. The values are plotted vesrus the TLS excitation probability $p$, Eq.~\eqref{eq-psi-pure}, with the red vertical line guiding the eye for the value $p=0.1$ and black arrows expressing the coherence and energy increase. The conditional increase of the final energy over the initial one, $(E_f-E_0)/E>0$, is achieved with maximum in the region of small $p$. The final entropy of coherence can be conditionally increased above the initial one in a region of small excitation probabilities $p$, up to the point $\tilde{C}_0\approx 0.96$. (\subref{fig-2-qub-eps}) Plot of the difference between the final and initial relative entropy of coherence,  $\Delta C=C_f-C_0$, Eq.~\eqref{eq-mixed-DeltaC}, for a pair of TLS in partially-coherent states with different values of $\epsilon$. The values are plotted versus the TLS excitation probability $p$, Eq.~\eqref{eq-mixed-2}, again the red vertical line guides the eye for the value $p=0.1$, as in the left panel. The final entropy of coherence can be conditionally increased above the initial one in a region of small excitation probabilities $p$, up to the point $\tilde{C}_0 (\epsilon)$, cf. Eq.~\eqref{eq-mixed-C0-bound}. The final and initial energy, $E_f/E$ and $E_0/E$ of the system is identical as for the case of pure states, cf. Fig.~\ref{fig-2-qub}, as energy is determined by diagonal terms of density matrix $\op{\rho}$ only and is independent of $\epsilon$.}
\end{figure}

The state \eqref{eq-pure-succ} determines the final energy $E_f$ and its variance $\langle \Delta E_f^2\rangle$ with respect to the Hamiltonian~\eqref{eq-Hamiltonian}, yielding
\begin{equation}
E_f=\bra{\Psi_f}\op{H}\ket{\Psi_f}=\frac{pE}{2-p},\quad \langle \Delta E_f^2\rangle=\frac{2p(1-p)E^2}{2-p},\quad p\neq 0.
\label{eq-pure-Ef}
\end{equation}
It can be compared to the initial average energy and the initial energy variance of the state $\eqref{eq-Psi-init}$
\begin{equation}
E_0=\bra{\Psi_i}\op{H}\ket{\Psi_i}=(2p-1)E,\quad \langle \Delta E_0^2\rangle=2p(1-p)E^2,
\label{eq-pure-Ei}
\end{equation}
yielding the energy gain
\begin{equation}
\Delta E=\frac{2(1-p)^2}{2-p}E.
\label{eq-pure-DE}
\end{equation}
The dependence of both on the excitation probability $p$, c.f. Eq.~\eqref{eq-psi-pure}, are shown in Fig.~\ref{fig-2-qub}. Clearly, sorting of the total ensemble of states $\ket{\Psi_i}$ increases the energy conditionally on the subensemble of states $\ket{\Psi_f}$, resulting in $\Delta E >0$, cf. Eq.~\eqref{eq-DeltaE}. We emphasize the fact, that $E_f\geq E_0$ for {\it all} excitation probabilities $p$ with the largest increase in the region $p\ll 1$, together with the decrease of the energy variance $\langle \Delta E_f^2\rangle < \langle \Delta E_0^2\rangle$ for all $0<p<1$. 
We can explain this effect by noting that the successful charging procedure cuts-off the lowest energy contribution $(-E)$, weighted by the ground state probability $(1-p)^2$. This amounts to new conditional populations of the energy eigenstates, Eq.~\eqref{eq-pure-succ}, the shift of the average energy $E_0\rightarrow E_f$, and lowering of the energy variance $\langle \Delta E_f^2\rangle < \langle \Delta E_0^2\rangle$, c.f. Eqs.~\eqref{eq-pure-Ef} and \eqref{eq-pure-Ei}, hence concentrating the energy distribution of the final battery. Together with the fact, that the average value of energy is increased by the protocol, $E_f>E_0$, one can understand the situation as an increase in the quality of coherent energy. Simultaneously, the projective measurement does not distinguish the states $\ket{e_1e_2},\ket{e_1g_2},\ket{g_1e_2}$ and therefore it allows not only to keep the initial coherence, but even to increase it.

Having characterized the initial and final states $\ket{\Psi_i}$, $\ket{\Psi_f}$, we can determine their coherence $C_0$, $C_f$, respectively, Eq.~\eqref{eq-rel-ent-coh}, in a straightforward way. The initial coherence of the state $\ket{\Psi_i}$ is (due to TLS independence) additive, and reads
\begin{equation}
C_0=-2\left[p\ln p+(1-p)\ln(1-p) \right],
\label{eq-pure-Ci}
\end{equation}
plotted in Fig.~\ref{fig-2-qub}. The analytic form of $C_f$ is, however, cumbersome, thus we present the result only graphically. It reveals that $C_f\geq C_0$ in the region of low excitation probabilities $p\leq p_0\approx 0.18$. The value $p_0$ is the solution of a transcendent equation $C_f(p)=C_0(p)$, not solvable analytically. Therefore, it is more informative to focus on the relation between the final and initial coherence and deal with $p$ as with an implicit parameter. Such an approach yields approximate functional relation valid for every $p$
\begin{equation}
C_f\approx \ln 2\left(\frac{2}{5}C_0+1\right).
\label{eq-pure-CfCi-relation}
\end{equation}
This result provides immediately $\lim_{C_0\rightarrow 0}C_f=\ln 2$, see Fig.~\ref{fig-2-qub}. The relation \eqref{eq-pure-CfCi-relation} captures the linear transformation ``amplifying'' $C_0\rightarrow C_f$ up to the boundary value $C_f=\tilde{C}_0\approx 0.96$. 
This represents the value of the initial/final coherence where the curves $C_f$ and $C_0$ cross each other. As such, $\tilde{C}_0$ represents the maximum value of the coherence attainable, if coherence {\it increase} is required. 

We would like to stress that our protocol basically decomposes any initial state Eq.~\eqref{eq-Psi-init} into component in the one dimensional $|g_1g_2\rangle$ subspace and its orthogonal complement from three dimensional subspace spanned by $\{|g_1e_2\rangle,|e_1g_2\rangle,|e_1e_2\rangle\}$. The validity of the values $\Delta C$ and $\Delta E$ directly stems from the validity of the form of post-measurement state in Eq.~\eqref{eq-pure-succ} {\it only} for $p\neq 0$. The initial state $|\Psi_i\rangle$, Eq.~\eqref{eq-Psi-init}, either has a non-zero component in the above mentioned 3D subspace (for $p>0$), or does not have it (for $p=0$) in which case there is no increase of the energy and coherence after the successful measurement, causing the discontinuity of the values at $p=0$. This fact is taken into account in all our derivations and is stressed by the condition $p\neq 0$ in Eq.~\eqref{eq-pure-succ} and the following.

\subsection*{Synthesizing the 2-TLS coherent battery made from mixed states}
\label{sub-2-mixed-states}

Whereas the previous subsection has introduced the main idea of our measurement-based charging process, the following lines will describe the effect of partial coherence of the TLS initial state. Such states are characterized, e.g., by the property ${\rm Tr}(\op{\rho}^2)<1$ and typically result from dephasing or spontaneous emission processes acting on the TLS. In the following lines, we will study these two archetypal processes individually. 

We start with the pure dephasing process acting individually on each TLS because it is usually the fastest process, which reduces coherence after it is harnessed from the environment. In this case, the pure initial state \eqref{eq-psi-pure} will be modified as 
\begin{eqnarray}
\op{\rho}_j=p\ket{e_j}\bra{e_j}+\epsilon\sqrt{p(1-p)}(\ket{e_j}\bra{g_j}+\ket{g_j}\bra{e_j})+(1-p)\ket{g_j}\bra{g_j},\quad j=1,2,
\label{eq-mixed-2}
\end{eqnarray}
where $0\leq\epsilon<1$. The total initial state of the system reads $\hat{\rho}_{i} =\hat{\rho}_1\otimes\hat{\rho}_2$, Eq.~\eqref{eq-mixed-2}.

The charging procedure is considered identical to the one in the previous subsection characterized by a pair of projectors \newline$\{\op{P}_0,\op{P}_1=\op{1}-\op{P}_0 \}$, Eq.~\eqref{eq-2-projector}, but applied to the state $\hat{\rho_{i}}$, resulting in the successfully charged state 
\begin{eqnarray}
\op{\rho}_f=\frac{\op{P}_1\op{\rho}_i\op{P}_1}{{p_s}},\quad p_s=p(2-p)\neq 0, \label{eq-mixed-succ}
\end{eqnarray} 
where $p_s$ is the same value of the success probability, Eq.~\eqref{eq-pure-succ}, independent of $\epsilon$.

We stress, that the partially coherent state, $\epsilon <1$, does not change the results for the average energy $E_f$, Eq.~\eqref{eq-E}, obtained in the previous subsection, cf. Eq.~\eqref{eq-pure-Ef} and Fig.~\ref{fig-2-qub}. This is intuitively the case, because $E_0$ as well as $E_f$, depend only on the diagonal terms appearing in the state~\eqref{eq-mixed-2}. Hence, to recall, the initial energy of the battery is
\begin{equation}
E_0= {\rm Tr}[\hat{\rho}_{i} \hat{H}]=(2p-1)E.
\label{eq-mixed-gibbs-tot-en}
\end{equation}

On the other hand $\epsilon <1$ strongly influences the values of the initial coherence $C_0=C(\hat{\rho}_i)=\sum_j C(\hat{\rho}_j)$, Eq.~\eqref{eq-mixed-2}, as well as the final one $C_f=C(\hat{\rho}_f)$, Eq.~\eqref{eq-mixed-succ}. Therefore, the mixedness of the initial state also affects the relation between $C_f$ and $C_0$, cf. Eq.~\eqref{eq-pure-CfCi-relation}, and their difference $\Delta C=C_f-C_0$
\begin{eqnarray}\label{eq-mixed-DeltaC}
\Delta C \approx \left(\frac{5-\epsilon}{10}\ln 2-1 \right)C_0+\epsilon^2\atanh{\epsilon^2}+\ln\sqrt{1-\epsilon^4}.
\end{eqnarray}
Plot of $\Delta C$, Eq.~\eqref{eq-mixed-DeltaC}, is shown in Fig.~\ref{fig-2-qub-eps} for several values of parameter $\epsilon$. Following the same logic as in the previous subsection, we characterize the upper bound on the initial coherence value $C_0$ that can be still increased. Equation~\eqref{eq-mixed-DeltaC} yields the following result
\begin{equation}
\Tilde{C}_0= \frac{\epsilon^2\atanh{\epsilon^2}+\ln\sqrt{1-\epsilon^4}}{1-\ln 2\left(5-\epsilon \right)/10}\approx \frac{\epsilon^4}{2\left[1-\ln 2\left(5-\epsilon \right)/10 \right]},
\label{eq-mixed-C0-bound}
\end{equation}
as the numerator of the middle fraction can be well approximated by the function $\epsilon^4/2$ up to $\epsilon\approx 0.8$. We see, Fig.~\ref{fig-2-qub-eps}, that for decreasing $\epsilon$ the maximum value $\Tilde{C}_0$ for which coherence increase is obtained, shifts to {\it lower} values of $p$. 

\begin{figure}[ht]
\begin{subfigure}{0.49\textwidth}
\begin{tikzpicture} 
  \node (img1)  {\includegraphics[width=.9\columnwidth]{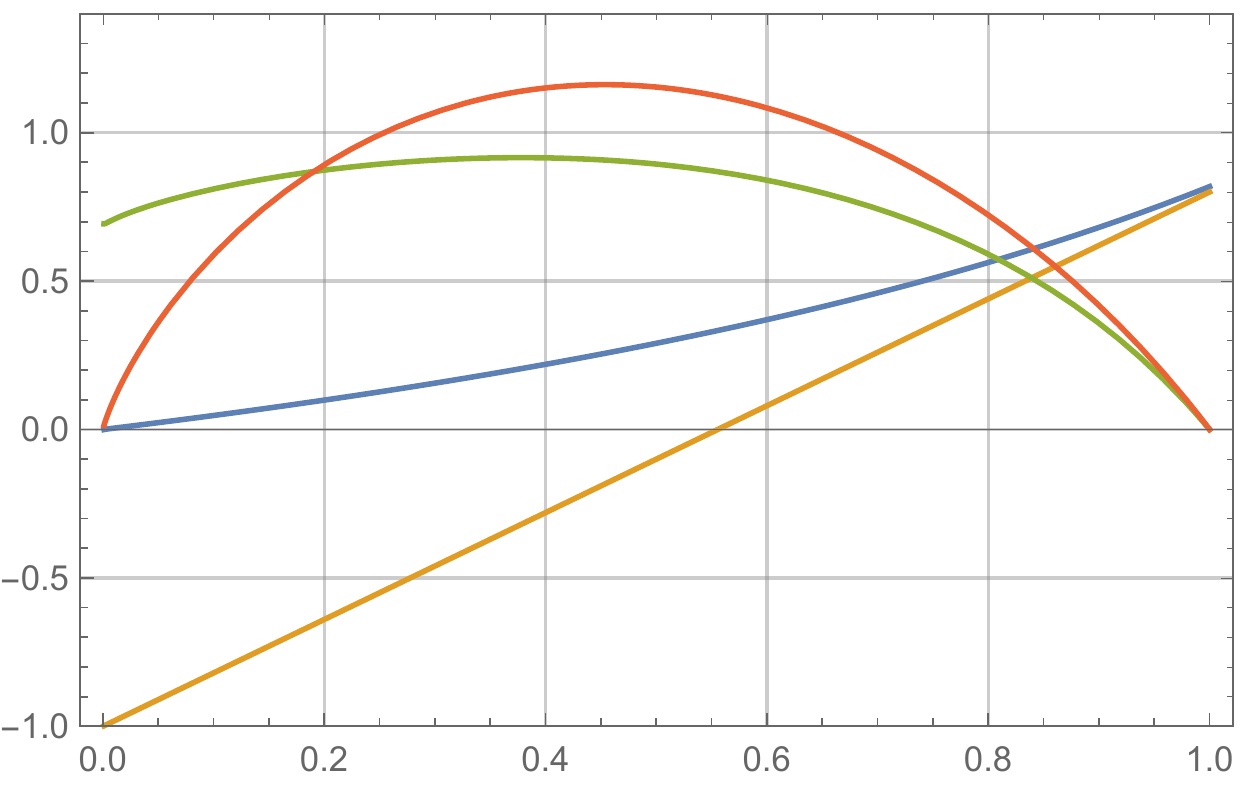}};
  \draw[red,thin] (-2.55,-2.1) -- (-2.55,2.4);
  \draw[black!30!blue,very thick,|->] (-2.55,.8) -- (-2.55,1.25);
  \draw[black!30!blue,very thick,|->] (-2.55,-1.75) -- (-2.55,-.1);
  \node[above=of img1, node distance=0cm, yshift=-6.5cm,xshift=0cm] {$p$};
  \node[above=of img1, node distance=0cm, yshift=-2.9cm,xshift=-.3cm] {{\color{black!30!blue}$\eta=0.1$}};
   \node[above=of img1, node distance=0cm, yshift=-3.cm,xshift=-1.8cm] {{\color{black!10!red}$\overline{C}_0$}};
   \node[above=of img1, node distance=0cm, yshift=-2.9cm,xshift=-3.0cm] {{\color{black!30!blue}$\Delta\overline{C}$}};
   \node[above=of img1, node distance=0cm, yshift=-4.9cm,xshift=-3.0cm] {{\color{black!30!blue}$\Delta\overline{E}$}};
   \node[above=of img1, node distance=0cm, yshift=-2.18cm,xshift=-2.2cm] {{\color{black!40!green}$\overline{C}_f$}};
    \node[above=of img1, node distance=0cm, yshift=-3.9cm,xshift=0.2cm] {{\color{black!30!blue}$\overline{E}_f/E$}};
    \node[above=of img1, node distance=0cm, yshift=-5.1cm,xshift=-0.4cm] {{\color{black!20!orange}$\overline{E}_0/E$}};
\end{tikzpicture}
\caption{}
\label{fig-mixed-spon-E}
\end{subfigure}
\hfill
\begin{subfigure}{0.49\textwidth}
\begin{tikzpicture} 
  \node (img1)  {\includegraphics[width=.9\columnwidth]{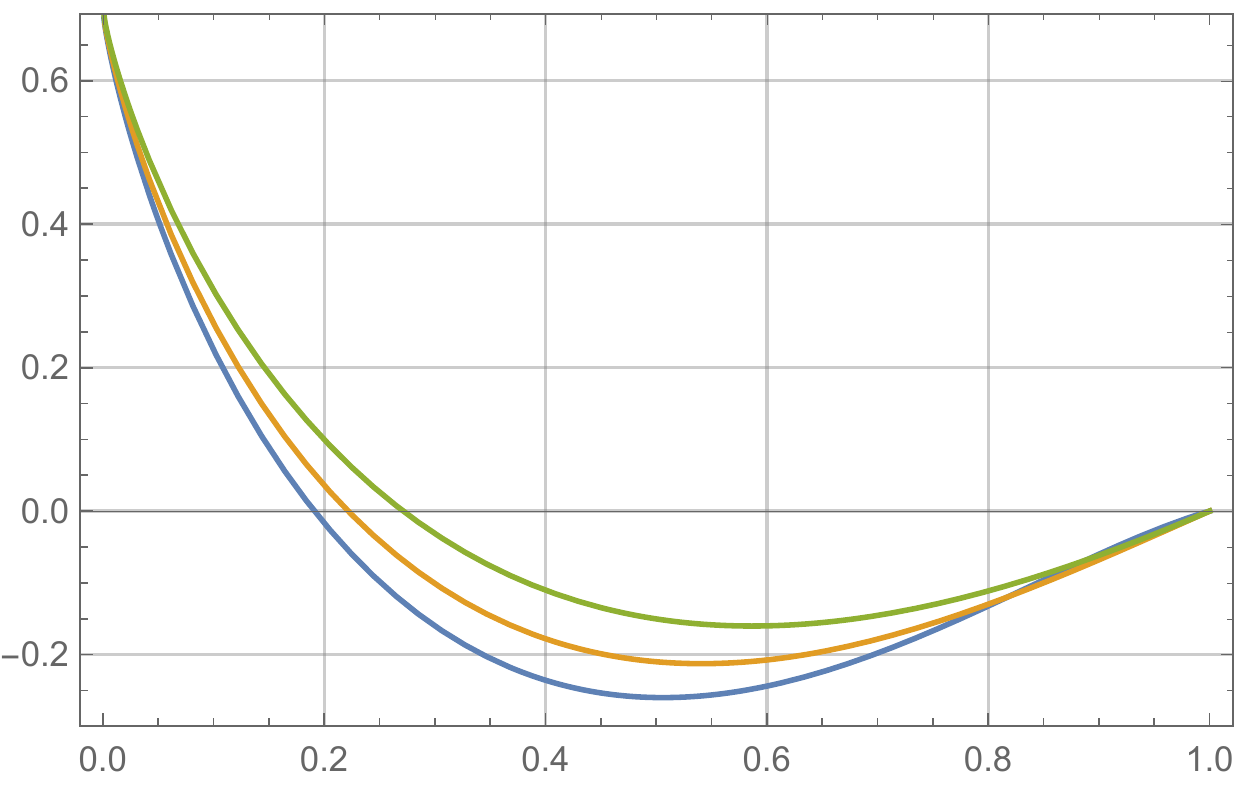}};
  \draw[red,thin] (-2.55,-2.1) -- (-2.55,2.4);
  \node[above=of img1, node distance=0cm, yshift=-6.5cm,xshift=0cm] {$p$};
   \node[above=of img1, node distance=0cm, yshift=-1.2cm,xshift=-3.5cm] {$\Delta \overline{C}$};
   \node[above=of img1, node distance=0cm, yshift=-3.5cm,xshift=-1.5cm] {{\color{black!50!green}$\eta=0.5$}};
    \node[above=of img1, node distance=0cm, yshift=-4.8cm,xshift=-0.2cm] {{\color{black!20!orange}$\eta=0.3$}};
    \node[above=of img1, node distance=0cm, yshift=-5.5cm,xshift=-1.8cm] {{\color{black!40!blue}$\eta=0.1$}};
\end{tikzpicture}
\caption{}
\label{fig-mixed-spont-eps}
\end{subfigure}
\caption{ (\subref{fig-mixed-spon-E}) The plot of the normalized average energy $\overline{E}_f/E$, $\overline{E}_0/E$ and the relative entropy of coherence $\overline{C}_f$, $\overline{C}_0$ for the final state $\overline{\op{\rho}}_f$, Eq.~\eqref{eq-mixed-spontaneous-succ}, and the initial state $\overline{\op{\rho}}_i$, Eq.~\eqref{eq-mixed-spontaneous},  with spontaneous emission probability $\eta=0.1$. The values are plotted versus the TLS excitation probability $p$, Eq.~\eqref{eq-psi-pure}, the red vertical line guides the eye for the value $p=0.1$, and the blue arrows express the coherence and energy increase. The conditional increase of the final energy over the initial one, $(\overline{E}_f-\overline{E}_0)/E>0$, is achieved with maximum in the region of small $p$. The final coherence can be conditionally increased above the initial one in a region 
with $\eta$ dependent boundary. (\subref{fig-mixed-spont-eps}) The plot of the difference between the final and initial coherence,  $\Delta \overline{C}=\overline{C}_f-\overline{C}_0$, Eq.~\eqref{eq-mixed-spontaneou-DeltaC}, for a pair of TLS in states after spontaneous emission with different values of $\eta$. For $\eta\ll 1$ one finds the increase in the region of small excitation probabilities $p$ (again the red vertical line guides the eye for the value $p=0.1$) up to the point $\tilde{C}_0$, Eq.~\eqref{eq-mixed-spontaneous-C0}. Remarkably, for increasing $\eta$, $\Delta \overline{C}$ increases as well, which comes at the expense of the corresponding probability of success $p_s$, Eq.~\eqref{eq-mixed-spontaneous-succ}. }
\end{figure}

Now we turn our attention to the initial state of the pair of TLS resulting from a process of spontaneous emission with probability $\eta$. Such a state has the form
\begin{eqnarray}\nonumber
\overline{\op{\rho}}_i&=&\left\{\left[(1-p)+\eta
p\right]\ket{g_1}\bra{g_1}+\sqrt{1-\eta}\sqrt{p(1-p)}(\ket{g_1}\bra{e_1}+\ket{e_1}\bra{g_1})+(1-\eta)p\ket{e_1}\bra{e_1}\right\}\\
&\otimes&\left\{\left[(1-p)+\eta
p\right]\ket{g_2}\bra{g_2}+\sqrt{1-\eta}\sqrt{p(1-p)}(\ket{g_2}\bra{e_2}+\ket{e_2}\bra{g_2})+(1-\eta)p\ket{e_2}\bra{e_2}\right\},\quad 0\leq\eta\leq 1.
\label{eq-mixed-spontaneous}
\end{eqnarray}
Subjecting this state to the projector based charging procedure $\{\op{P}_0,\op{P}_1=\op{1}-\op{P}_0 \}$, Eq.~\eqref{eq-2-projector}, yields successful post measurement state
\begin{eqnarray}
\overline{\op{\rho}}_f=\frac{\op{P}_1\overline{\op{\rho}}_i\op{P}_1}{{\overline{p}_s}},\quad\overline{p}_s=(1-\eta)p[2-p(1-\eta)]\neq 0. \label{eq-mixed-spontaneous-succ}
\end{eqnarray} 
The initial energy of the battery now reads
\begin{eqnarray}
\overline{E}_0=E\left[2p(1-\eta)-1\right]. 
\label{eq-mixed-spontaneous-E0}
\end{eqnarray}
As in the previous lines it is more useful to show the approximate result for coherence change $\Delta C$, Eq.~\eqref{eq-DeltaC}, than the individual summands, yielding
\begin{eqnarray}
\Delta \overline{C}\approx \left[\frac{2\ln{2}}{5}\left(1+\frac{5}{2}\eta \right)-1 \right]\overline{C}_0+\ln{2},
\label{eq-mixed-spontaneou-DeltaC}
\end{eqnarray}
a good approximation for $0\leq\eta\leq 1/3$. The interesting fact that $\Delta \overline{C}$ increases with $\eta$ stems from the different speed at which $\overline{C}_0$ and $\overline{C}_f$ decrease with $\eta$. This opens the gap $\Delta \overline{C}$ between them and simultaneously increases the lower-value crossing point, $C_f=C_0$, in which the charging protocol stops increasing the coherence. This value reads (if  such point of coincidence exists), see Fig.~\ref{fig-mixed-spont-eps},
\begin{eqnarray}
\Tilde{C}_0\approx \ln{2}\left[1-\frac{2\ln{2}}{5}\left(1+\frac{5}{2}\eta \right) \right]^{-1}.
\label{eq-mixed-spontaneous-C0}
\end{eqnarray}
Comparison of Eqs.~\eqref{eq-mixed-spontaneous-C0}~and~\eqref{eq-mixed-C0-bound} reveals one striking difference. Although both processes (dephasing and spontaneous emission) of the initial states decrease the initial value of coherence $C_0$ in a similar way, the effect of charging yields results with qualitatively opposite behavior. Whereas lowering $\epsilon$, Eq.~\eqref{eq-mixed-2}, quickly {\it decreases} $\Tilde{C}_0$ (Fig.~\ref{fig-2-qub-eps}), the increase of $\eta$, Eq.~\eqref{eq-mixed-spontaneous}, leads to the {\it increase} of $\Tilde{C}_0$ (Fig.~\ref{fig-mixed-spont-eps}).

\subsection*{Use of POVM for optimizing $\Delta E$ or $\Delta C$ at the expense of success probability }
\label{sub-POVM}

This subsection describes a generalization of the projection-based ideas introduced in the previous subsections. As was presented, the projections are capable of increasing both the energy and coherence by some amount, captured by the results for $\Delta E>0$, Eq.~\eqref{eq-pure-DE}, and $\Delta C>0$, Eq.~\eqref{eq-mixed-DeltaC}. It is known that generalized measurements, positive-operator-valued-measures (POVMs), can be useful if a compromise between incompatible features has to be found. The POVM is mathematically represented by operators (elements) with the property $\op{M}\op{M}\neq\op{M}$, valid for any element of POVM. Due to fundamentally different properties of POVM compared to the projection, we can expect improvements in the behavior of $\Delta E$ and $\Delta C$. To check this expectation, we numerically generalize and optimize the charging protocol for the use of generalized measurement elements, c.f. Eq.~\eqref{eq-2-projector}, of the form $\lbrace \op{M}_0,\op{M}_1=\op{1}-\op{M}_0 \rbrace$ with 
\begin{equation}
\op{M}_0=a\ket{g_1g_2}\bra{g_1g_2}+b\,(\ket{g_1e_2}\bra{g_1e_2}+\ket{e_1g_2}\bra{e_1g_2}),\quad 0\leq a,b\leq 1,
\label{eq-POVM}
\end{equation}
applied to the pure initial state $\ket{\Psi_i}$, Eq.~\eqref{eq-Psi-init}. Clearly, the case $a=1$, $b=0$ represents the projector used at the beginning of this section. On the other hand, $a=1$, $b=1$  is clearly an optimal choice for maximization of the final energy, as it results in removing all the populations except those of the state $\ket{e_1e_2}$, preparing the analogy of the successful state \eqref{eq-pure-succ} with maximum energy, but certainly zero coherence in the energy eigenbasis. Hence, equation~\eqref{eq-POVM} represent a generalization of the projections to the set of POVM's, but is still compatible with the previous important choice of measurement elements that are diagonal in the eigenbasis of the Hamiltonian \eqref{eq-Hamiltonian}. 

\begin{figure}[ht]
\begin{subfigure}{0.49\textwidth}
\begin{tikzpicture} 
  \node (img1)  {\includegraphics[width=.9\linewidth]{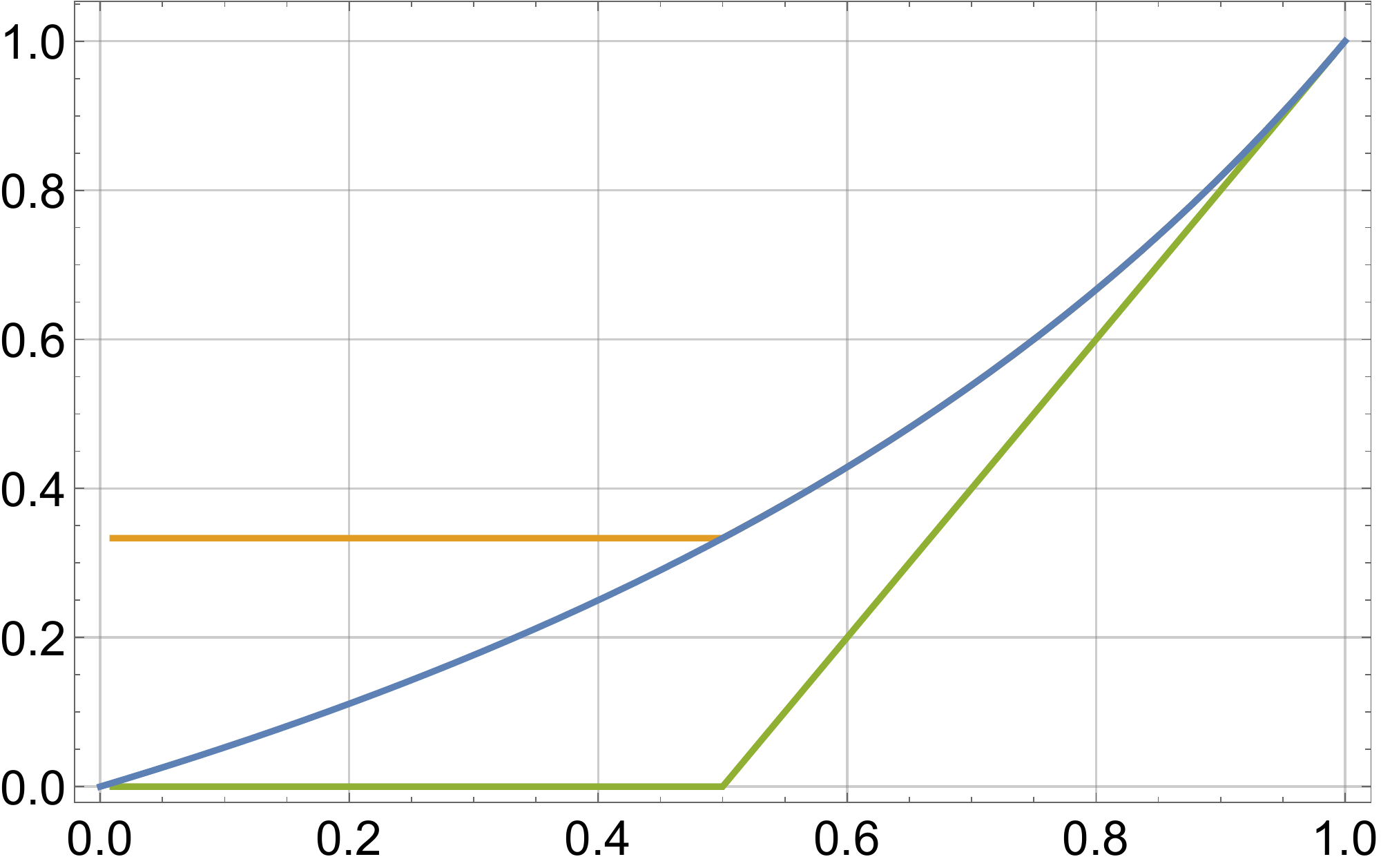}};
  \draw[red,thin] (-1.9,-2.1) -- (-1.9,2.5);
  \draw[black!20!orange,very thick,|->] (-1.9,-1.5) -- (-1.9,-0.6);
  \draw[black!60!green,very thick,|->] (-1.9,-1.5) -- (-1.9,-2);
  \node[above=of img1, node distance=0cm, yshift=-6.3cm,xshift=0cm] {$p$};
    \node[above=of img1, node distance=0cm, yshift=-5.4cm,xshift=-0.9cm] {{\color{blue}$E_f/E$}};
    \node[above=of img1, node distance=0cm, yshift=-4.1cm,xshift=-.9cm] {{\color{black!20!orange}$E_f(\overline{b})/E$}};
    \node[above=of img1, node distance=0cm, yshift=-5.1cm,xshift=1.8cm] {{\color{black!60!green}$E_f(\overline{a},\overline{b})/E$}};
\end{tikzpicture}
\caption{\label{fig-2-qub-povm-en}}
\end{subfigure}
\begin{subfigure}{0.49\textwidth}
\begin{tikzpicture} 
  \node (img1)  {\includegraphics[width=.9\linewidth]{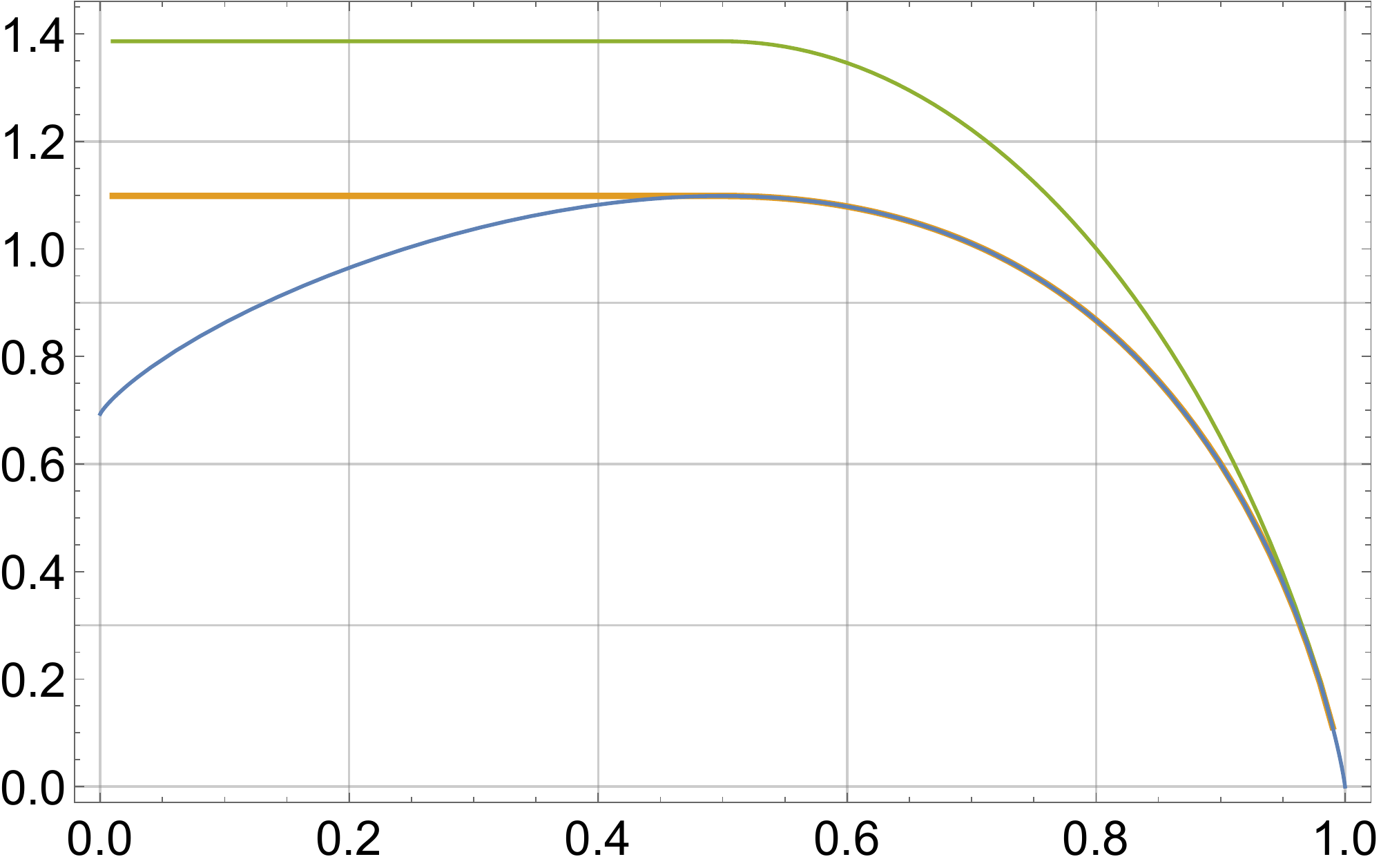}};
  \draw[red,thin] (-1.9,-2.1) -- (-1.9,2.5);
  \draw[black!60!green,very thick,|->] (-2,.9) -- (-2,2.2);
  \draw[black!20!orange,very thick,|->] (-1.8,0.9) -- (-1.8,1.35);
  \node[above=of img1, node distance=0cm, yshift=-6.3cm,xshift=0cm] {$p$};
    \node[above=of img1, node distance=0cm, yshift=-3.6cm,xshift=-2.5cm] {{\color{blue}$C_f$}};
    \node[above=of img1, node distance=0cm, yshift=-2.2cm,xshift=-2.8cm] {{\color{black!20!orange}$C_f(\overline{b})$}};
    \node[above=of img1, node distance=0cm, yshift=-2cm,xshift=-1.0cm] {{\color{black!60!green}$C_f(\overline{a},\overline{b})$}};
\end{tikzpicture}
\caption{\label{fig-2-qub-povm-c}}
\end{subfigure}
\caption{Plots of the optimal energy after the charging (\subref{fig-2-qub-povm-en}) and coherence (\subref{fig-2-qub-povm-c}) for general POVM (green), cf. Eq.~\eqref{eq-POVM}, and its restricted class with $a=1$ (orange), applied to the initial state $\ket{\Psi_i}$, Eq.~\eqref{eq-Psi-init}. These are compared to blue lines showing $E_f/E$ and $C_f$ from Fig.~\ref{fig-2-qub}, respectively. The energy values are given for optimal values of parameters $\overline{a}$ and $\overline{b}$ for which the final coherence $C_f(\overline{b})$ and  $C_f(\overline{a},\overline{b})$ is maximized. The coherence $C_f(\overline{a},\overline{b})$ (green line), for the general type of POVM, shows substantial increase over the whole range of $p$ (e.g., green arrow for $p=0.2$ in Fig.~\ref{fig-2-qub-povm-c}), compensated by  smaller increase in the energy compared to the simple projection, Eq.~\eqref{eq-2-projector} (e.g., green arrow for $p=0.2$ in Fig.~\ref{fig-2-qub-povm-en}). The results for restricted type of POVM (orange lines) show improvement of the coherence values (e.g., orange arrow for $p=0.2$ in Fig.~\ref{fig-2-qub-povm-c}) over the interval of $p<1/2$, and simultaneous increase of the post measurement energy values (e.g., orange arrow for $p=0.2$ in Fig.~\ref{fig-2-qub-povm-en}) on the same interval of $p$, with respect to the simple projector case, Eq.~\eqref{eq-2-projector}.}
\label{fig-2-qub-povm}
\end{figure}
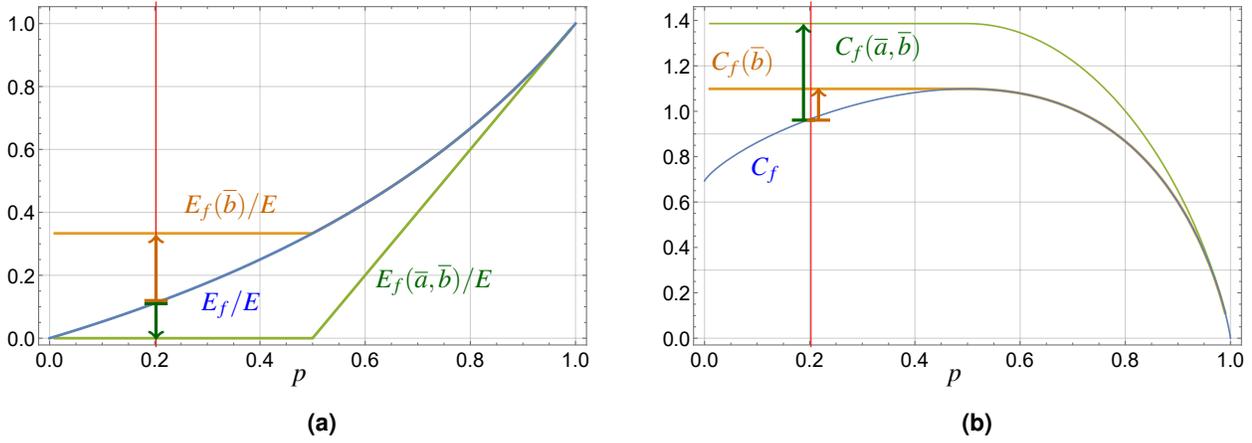

Due to the more complicated structure of the analytical results obtained using the elements of \eqref{eq-POVM}, we restrict ourselves to a graphical representation of the results, see Fig.~\ref{fig-2-qub-povm}. They present values of the coherence optimized for each fixed excitation probability $p$, Eq.~\eqref{eq-psi-pure}, where the optimization is done over a certain range of $a$, $b$ parameters, see below. The energy plots show, for each fixed $p$, the post-measurement value corresponding to the optimal $\overline{a}$ (or $\overline{b}$) parameters determined by optimization of the coherence.

Figure~\ref{fig-2-qub-povm} shows the results for a general class of POVM, cf. Eq.~\eqref{eq-POVM}, and its restricted analogue with $a=1$, $0\leq b\leq 1$. We have chosen these two classes, because they represent POVM yielding two qualitatively different results, and we compare them both with the projector case $a=1$, $b=0$, see Fig.~\ref{fig-2-qub-povm}. The results for the more general class indicate a substantial coherence increase over the whole range of $p$, reaching the maximum attainable value $C_f(\overline{a}, \overline{b})\approx \ln{4}$  for a pair of TLS, but this is paid by a smaller increase in the energy compared to the simple projection, Eq.~\eqref{eq-2-projector} and Fig.~\ref{fig-2-qub-povm}. For the latter POVM class, we found the increase of coherence in the lower $p$ region (up to $p=1/2$) reaching $C_{f}(\overline{b})=\ln{3}<\ln{4}$, Fig.~\ref{fig-2-qub-povm-c}, and higher increase in energy compared to the simple projection, Eq.~\eqref{eq-2-projector}, whereas POVM defined in Eq.~\eqref{eq-POVM} gives an increase in energy with respect to the initial value $E_0$ only in the range of $0\leq p \leq 1/2$. This increase is significantly lower compared to the projector-based protocol, Fig.~\ref{fig-2-qub-povm-en}.

The optimization process for both classes of POVMs was made with respect to the final post measurement coherence $C_{f}$,  Eq.~\eqref{eq-rel-ent-coh}. Therefore, the general class, Eq.~\eqref{eq-POVM}, was optimized over $a$ and $b$ for each fixed $p$, resulting into $C_{f}(\overline{a}, \overline{b})$. Figure~\ref{fig-2-qub-povm-en} shows the corresponding final energy $E_{f}(\overline{a}, \overline{b})/E$. Similarly, the second class of POVM, restricted with $a=1$, was optimized over a single parameter $b$, where $C_{f}(\overline{b})$ is the corresponding final coherence with optimized values $\overline{b}$. 

The relation between parameters $\overline{a}$ and $\overline{b}$, which maximize coherence $C_f(\overline{a},\overline{b})$, Eq.~\eqref{eq-POVM}, has the form $\overline{b}=1-\sqrt{1-\overline{a}}$. Hence, we can rewrite the coherence-maximizing POVM, Eq.~\eqref{eq-POVM}, as 
\begin{equation}
\op{K}_0=\overline{a}\ket{g_1g_2}\bra{g_1g_2}+(1-\sqrt{1-\overline{a}})\,(\ket{g_1e_2}\bra{g_1e_2}+\ket{e_1g_2}\bra{e_1g_2}),\quad 0\leq \overline{a}\leq 1,
\label{eq-POVM-1}
\end{equation}
whereas the relationship between the optimal value $\overline{a}$ and the excitation probability $p$ can be well approximated by the polynomial dependence $\overline{a}\approx 1-3/2p^2-8p^4$ in the interval $0\leq p\leq 2/5$.


To summarize this subsection, we found that due to the presence of additional free parameters $a$, $b$, Eq.~\eqref{eq-POVM}, which allows for subsequent optimization, POVM protocol offers more resulting options compared to the simple projector based protocol. We have presented two different types of results, out of a number of options. POVM allows both for increase of $C_f(\overline{a},\overline{b})$ above the projector value $C_f$, together with simultaneous decrease of $E_f(\overline{a},\overline{b})$ below $E_f$, see Fig.~\ref{fig-2-qub-povm}. On the other hand, different class of POVM offers the possibility of increasing both $E_f(\overline{b})$ and $C_f(\overline{b})$ over the corresponding projector values, see Fig.~\ref{fig-2-qub-povm}.
We would like to note, that although POVMs have a similar positive effect (increased average energy and coherence) as the projectors, they may on the other hand increase the variance of energy. This fact underlines the qualitatively different nature of POVMs compared to projection measurements.

\section*{Synthesizing $N$-TLS coherent battery made from pure states}
\label{sec-N-systems}

This section generalizes the results described for a pair of TLS in the previous section. Here, a larger number of systems is subjected to the charging protocol. Namely, we consider a battery given by $N\geq 2$ noninteracting copies of TLS, characterized by the total Hamiltonian
\begin{equation}
\label{eq-Ham-N}
\hat{H}^{(N)}= \sum_{j=1}^{N} \op{H}_{j},
\end{equation} 
where $\op{H}_{j}$ are respective Hamiltonians of the $j$-th TLS, given in Eq.~\eqref{eq-Hams}.
In the following we restrict our attention only to the pure ($\epsilon =1$) initial states of the system, cf. Eq.~\eqref{eq-mixed-2}), noting that generalization can be done numerically in a straightforward way.

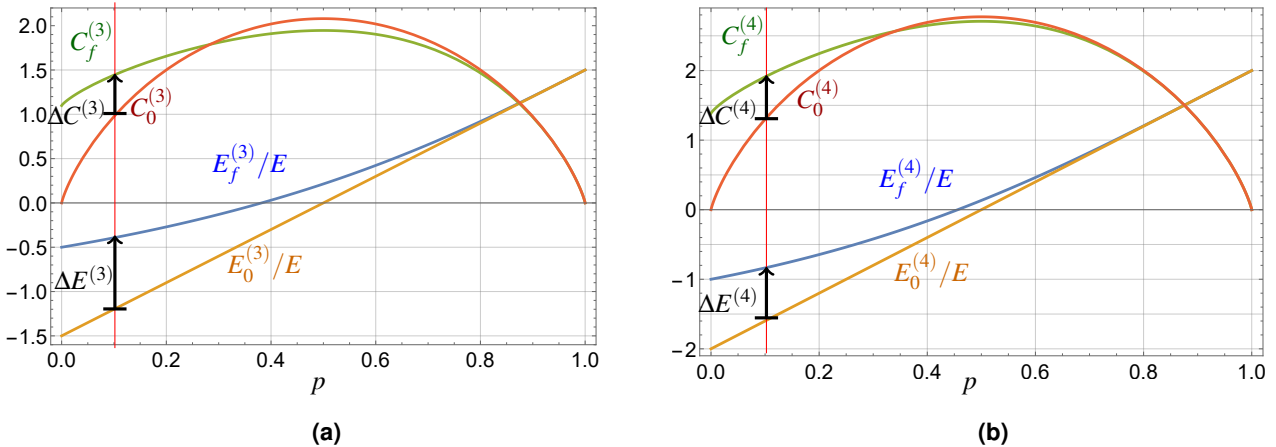
\begin{figure}[ht]
\begin{subfigure}{0.5\textwidth}
\begin{tikzpicture} 
  \node (img1)  {\includegraphics[width=.9\linewidth]{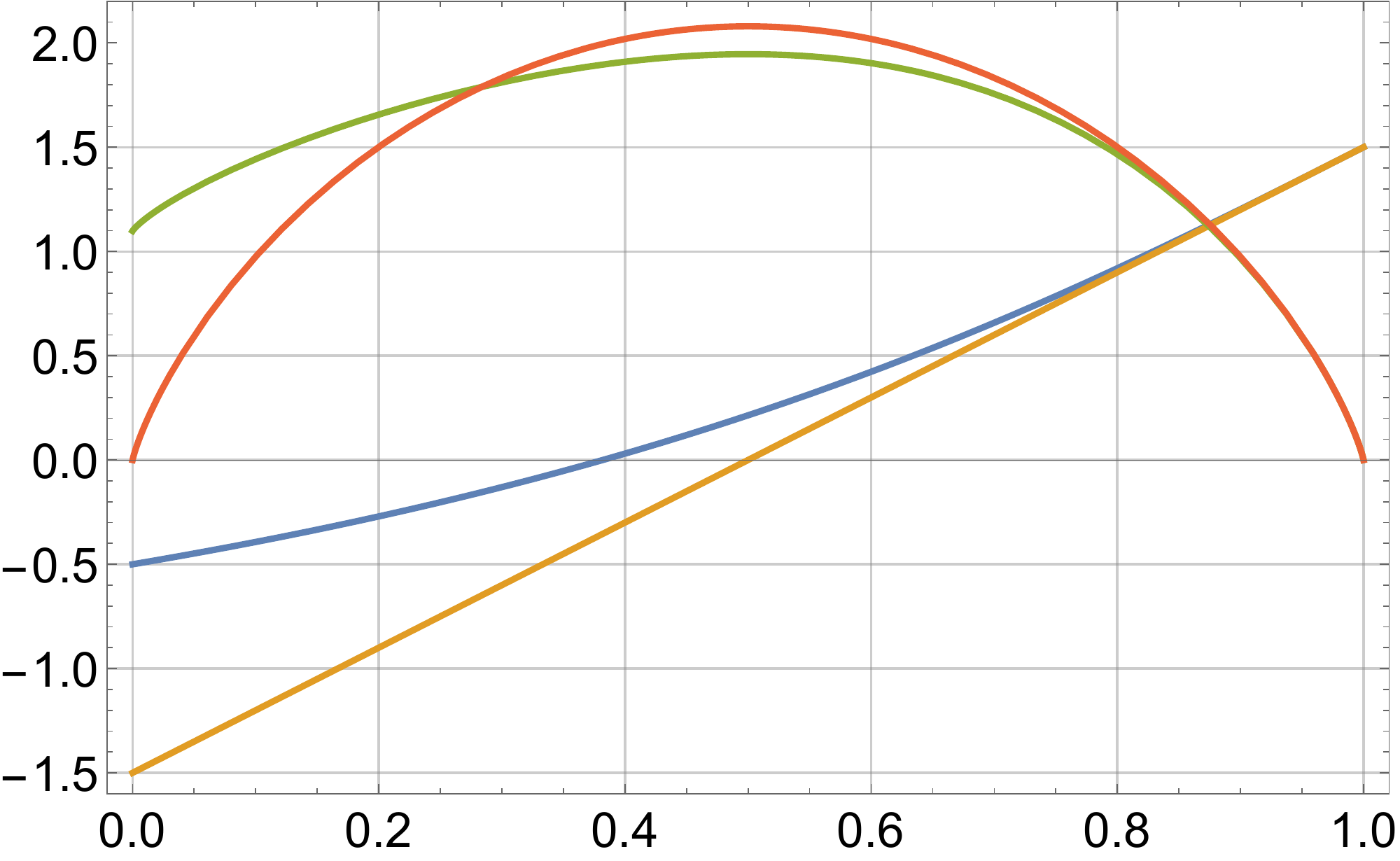}};
  \draw[red,thin] (-2.5,-2.1) -- (-2.5,2.5);
  \draw[black,very thick,|->] (-2.5,1.0) -- (-2.5,1.55);
  \draw[black,very thick,|->] (-2.5,-1.6) -- (-2.5,-.6);
  \node[above=of img1, node distance=0cm, yshift=-2.8cm,xshift=-3.0cm] {{\color{black}$\Delta C^{(3)}$}};
  \node[above=of img1, node distance=0cm, yshift=-5.0cm,xshift=-2.9cm] {{\color{black}$\Delta E^{(3)}$}};
  \node[above=of img1, node distance=0cm, yshift=-6.4cm,xshift=.2cm] {$p$};
   \node[above=of img1, node distance=0cm, yshift=-2.8cm,xshift=-2.cm] {{\color{black!40!red}$C_0^{(3)}$}};
   \node[above=of img1, node distance=0cm, yshift=-1.98cm,xshift=-2.8cm] {{\color{black!60!green}$C_f^{(3)}$}};
    \node[above=of img1, node distance=0cm, yshift=-3.6cm,xshift=-0.7cm] {{\color{blue}$E_f^{(3)}/E$}};
    \node[above=of img1, node distance=0cm, yshift=-4.9cm,xshift=-0.5cm] {{\color{black!20!orange}$E_0^{(3)}/E$}};
\end{tikzpicture}
\caption{\label{fig-3-qub}}
\end{subfigure}
\hfill
\begin{subfigure}{0.5\textwidth}
\begin{tikzpicture} 
  \node (img1)  {\includegraphics[width=.9\linewidth]{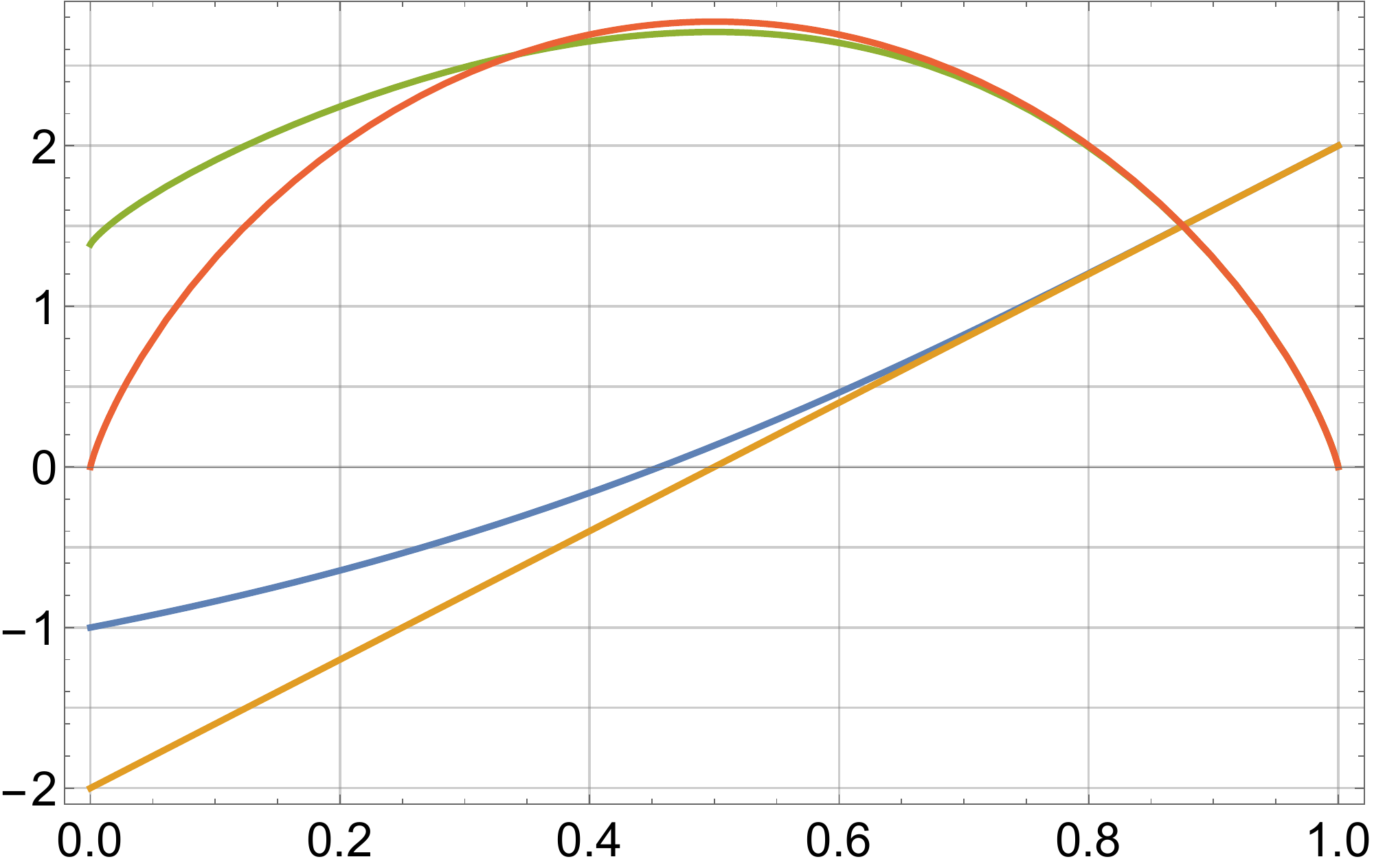}};
  \draw[red,thin] (-2.7,-2.1) -- (-2.7,2.5);
  \draw[black,very thick,|->] (-2.7,1.0) -- (-2.7,1.6);
  \draw[black,very thick,|->] (-2.7,-1.65) -- (-2.7,-.95);
  \node[above=of img1, node distance=0cm, yshift=-2.8cm,xshift=-3.2cm] {{\color{black}$\Delta C^{(4)}$}};
  \node[above=of img1, node distance=0cm, yshift=-5.3cm,xshift=-3.2cm] {{\color{black}$\Delta E^{(4)}$}};
  \node[above=of img1, node distance=0cm, yshift=-6.4cm,xshift=0cm] {$p$};
   \node[above=of img1, node distance=0cm, yshift=-2.7cm,xshift=-2.cm] {{\color{black!40!red}$C_0^{(4)}$}};
   \node[above=of img1, node distance=0cm, yshift=-1.9cm,xshift=-3.0cm] {{\color{black!60!green}$C_f^{(4)}$}};
   \node[above=of img1, node distance=0cm, yshift=-3.8cm,xshift=-0.7cm] {{\color{blue}$E_f^{(4)}/E$}};
    \node[above=of img1, node distance=0cm, yshift=-5cm,xshift=-0.5cm] {{\color{black!20!orange}$E_0^{(4)}/E$}};
\end{tikzpicture}
\caption{\label{fig-4-qub}}
\end{subfigure}
\caption{The plot of the normalized average energy $E_f^{(N)}/E$ \eqref{eq-pure-Ef-N}, $E_0^{(N)}/E$ \eqref{eq-N-Ei}, and the relative entropy of coherence $C_f^{(N)}$ \eqref{eq-N-coh-pure-final}, $C_0^{(N)}$ \eqref{eq-N-coh-pure-init}, for the final state $\ket{\Psi_{f}^{(N)}}$, Eq.~\eqref{eq-pure-succ-N}, and the initial state $\ket{\Psi_{i}^{(N)}}$, Eq.~\eqref{eq-N-Psii}, for $N=3$ (\subref{fig-3-qub}) and $N=4$ (\subref{fig-4-qub}). The values are plotted versus the higher level excitation probability $p$,  Eq.~\eqref{eq-psi-pure}. The average energy gain, $0\leq E_f^{(N)}-E_0^{(N)}$, is achieved with maximum for $p\rightarrow 0$. The final entropy of coherence $C_f^{(N)}$ can be conditionally increased above the initial one $C_0^{(N)}$ in a region of small excitation probabilities $p$. The gain of coherence $\Delta C ^{(4)} >\Delta C ^{(3)}>\Delta C ^{(2)}$ increases with the number $N$ of TLS constituting the battery, whereas the energy gain is limited by $E_f^{(N)}-E_0^{(N)}\leq E$ for every $N$, see Eq.~\eqref{eq-pure-DE-N}. The red vertical line is a guide for the eye at $p=0.1$ and the black arrows indicate the energy and coherence increase for each respective $N$.}
\label{fig-N-qub-global}
\end{figure}

The total initial state of the $N$ TLS's battery is therefore, cf. Eq.~\eqref{eq-psi-pure},
\begin{equation}
\ket{\Psi_{i}^{(N)}}=\displaystyle{\bigotimes_{j=1}^{N}\ket{\psi_j}},
\label{eq-N-Psii}
\end{equation}
where $j=1,\cdots,N$ labels the copies of TLS. Subsequently, the initial energy of the system is given by a straightforward (due to its additivity) generalization of Eq.~\eqref{eq-pure-Ei}, yielding
\begin{equation}
E_{0}^{(N)}=\frac{N E_{0}}{2}=\frac{NE}{2}(2p-1).\label{eq-N-Ei}
\end{equation}

The charging process is analogous to previous subsections and consists of application of a pair of global projectors \newline$\{\op{P}_0^{(N)},\; \op{P}_1^{(N)}=\op{1}-\op{P}_0^{(N)} \}$, where 
\begin{equation}
\op{P}_0^{(N)}=\displaystyle{\bigotimes_{j=1}^{N}\ket{g_j}\bra{g_j}},\quad \op{P}_1^{(N)}=\op{1}-\op{P}_0^{(N)},
\end{equation}
is the projector on the global ground state of the system, cf. Eq.~\eqref{eq-2-projector}. 
Application of the projector $\op{P}_1^{(N)}$ to the state \eqref{eq-N-Psii} results in successfully charged state 
\begin{equation}
\ket{\Psi_{f}^{(N)}}=\frac{\op{P}_1^{(N)}\ket{\Psi_i^{(N)}}}{\sqrt{p_s^{(N)}}},\quad p_s^{(N)}=1-(1-p)^N \neq 0, 
\label{eq-pure-succ-N}
\end{equation}
with the probability of success $p_s^{(N)}=1-(1-p)^N$ converging to $1$ with $N$, for any finite $p\neq 0$, and the corresponding energy of the final, successful state
\begin{equation}
E_{f}^{(N)}=\frac{NE}{2}\left[\frac{2p}{1-(1-p)^N}-1\right].
\label{eq-pure-Ef-N}
\end{equation} 
From these results we obtain an upper bound on the energy increase 
\begin{eqnarray}
\Delta E^{(N)}\equiv E_{f}^{(N)}-E_{0}^{(N)}=E\left[\frac{Np(1-p)^N}{1-(1-p)^N} \right]\approx E(1-Np)\leq E,\quad p\ll 1,\,N\gg 1.
\label{eq-pure-DE-N}
\end{eqnarray}

The final energy $E_{f}^{(N)}$, Eq.~\eqref{eq-pure-Ef-N}, after the charging process as a function of $p$, parametrized by the number $N$ of TLS, is plotted in Fig.~\ref{fig-N-qub-global}, where we present the results for pure initial states of $N=3$ and $N=4$ TLS, see Fig.~\ref{fig-3-qub} and Fig.~\ref{fig-4-qub}, respectively. By comparison to results for a pair of TLS, Fig.~\ref{fig-2-qub}, we can see the behavior of the final energy $E_{f}^{(N)}$ with the growth of $N$. The final energy in the case of successful charging is increased for all $p$, however, the energy gain reads $\Delta E^{(N)} \leq E$, and it is thus upper bounded by the energy gap $E$ for all $N$. Thus, the goal of the charging protocol (energy increase) is achieved, although the change of the energy {\it relative} to the total initial energy available, Eq.~\eqref{eq-N-Ei}, is decreasing as $|\Delta E^{(N)}/E_0^{(N)}|\propto N^{-1}$ in the range $p\ll 1$, where the energy increase is maximum.

The expression for the final coherence $C_f^{(N)}$ for the successful measurement outcome in the case of pure initial state \eqref{eq-N-Psii} (the term corresponding to the von Neumann entropy vanishes) reads
\begin{eqnarray}
C_f^{(N)}=-\sum_{k=1}^{N}\binom{N}{k}\frac{p^k(1-p)^{N-k}}{1-(1-p)^{N}}\ln\left[{\frac{p^k(1-p)^{N-k}}{1-(1-p)^{N}}}\right],\quad \lim_{p\rightarrow 0}C_f^{(N)}=\ln{N},
\label{eq-N-coh-pure-final}
\end{eqnarray}
and should be compared to the coherence of the initial pure state, Eq.~\eqref{eq-N-Psii}, 
\begin{eqnarray}
C_0^{(N)}=-\sum_{k=0}^{N}\binom{N}{k}p^k(1-p)^{N-k}\ln\left[{p^k(1-p)^{N-k}}\right].
\label{eq-N-coh-pure-init}
\end{eqnarray}

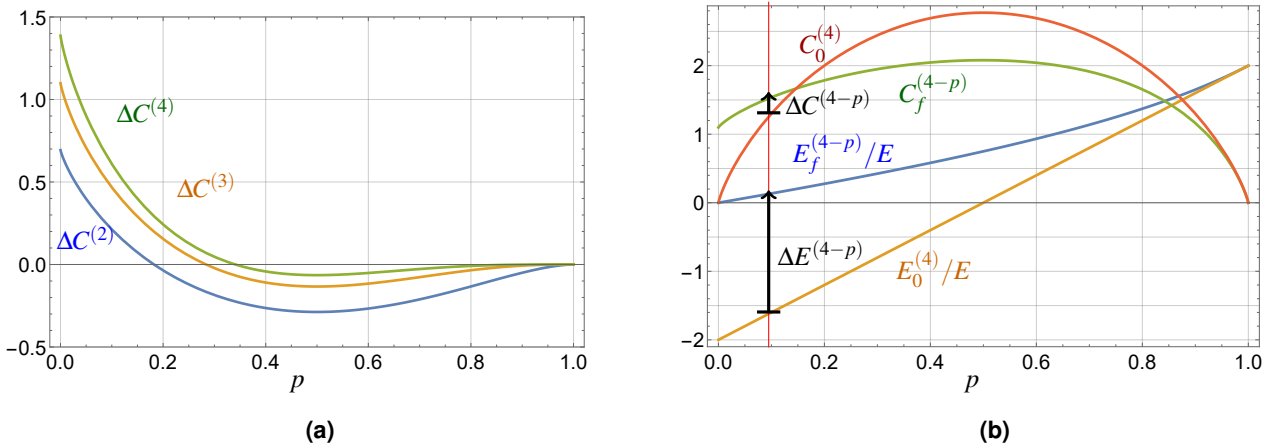
\begin{figure}[ht]
\begin{subfigure}{0.49\textwidth}
\begin{tikzpicture} 
  \node (img1)  {\includegraphics[width=.9\linewidth]{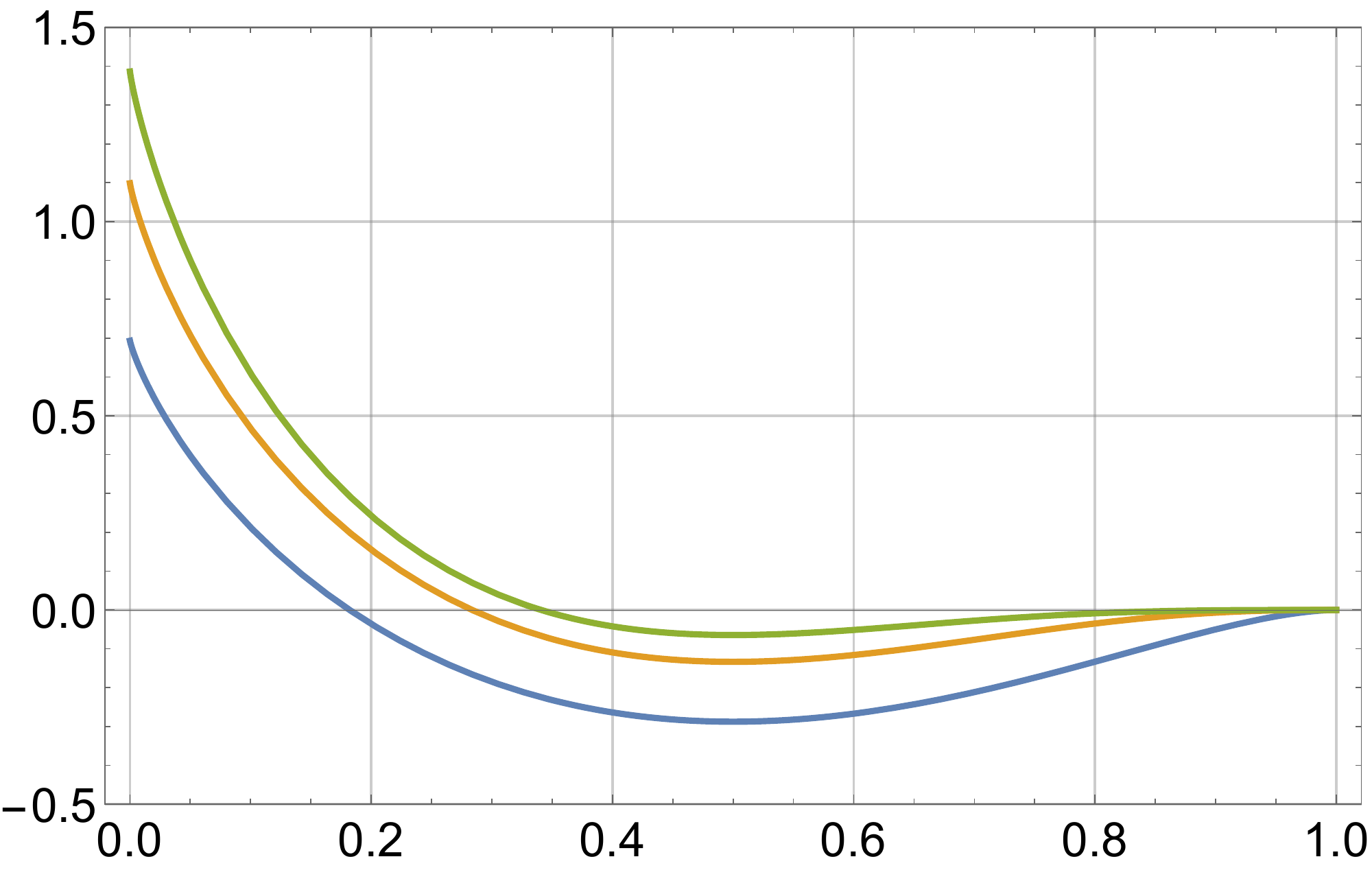}};
  \node[above=of img1, node distance=0cm, yshift=-4.5cm,xshift=-2.8cm] {{\color{blue}$\Delta C^{(2)}$}};
  \node[above=of img1, node distance=0cm, yshift=-3.8cm,xshift=-1.2cm] {{\color{black!20!orange}$\Delta C^{(3)}$}};
  \node[above=of img1, node distance=0cm, yshift=-6.4cm,xshift=-.0cm] {$p$};
   \node[above=of img1, node distance=0cm, yshift=-2.8cm,xshift=-2.cm] {{\color{black!60!green}$\Delta C^{(4)}$}};
\end{tikzpicture}
\caption{\label{fig-coh}}
\end{subfigure}
\hfill
\begin{subfigure}{0.49\textwidth}
\begin{tikzpicture} 
  \node (img1)  {\includegraphics[width=.9\linewidth]{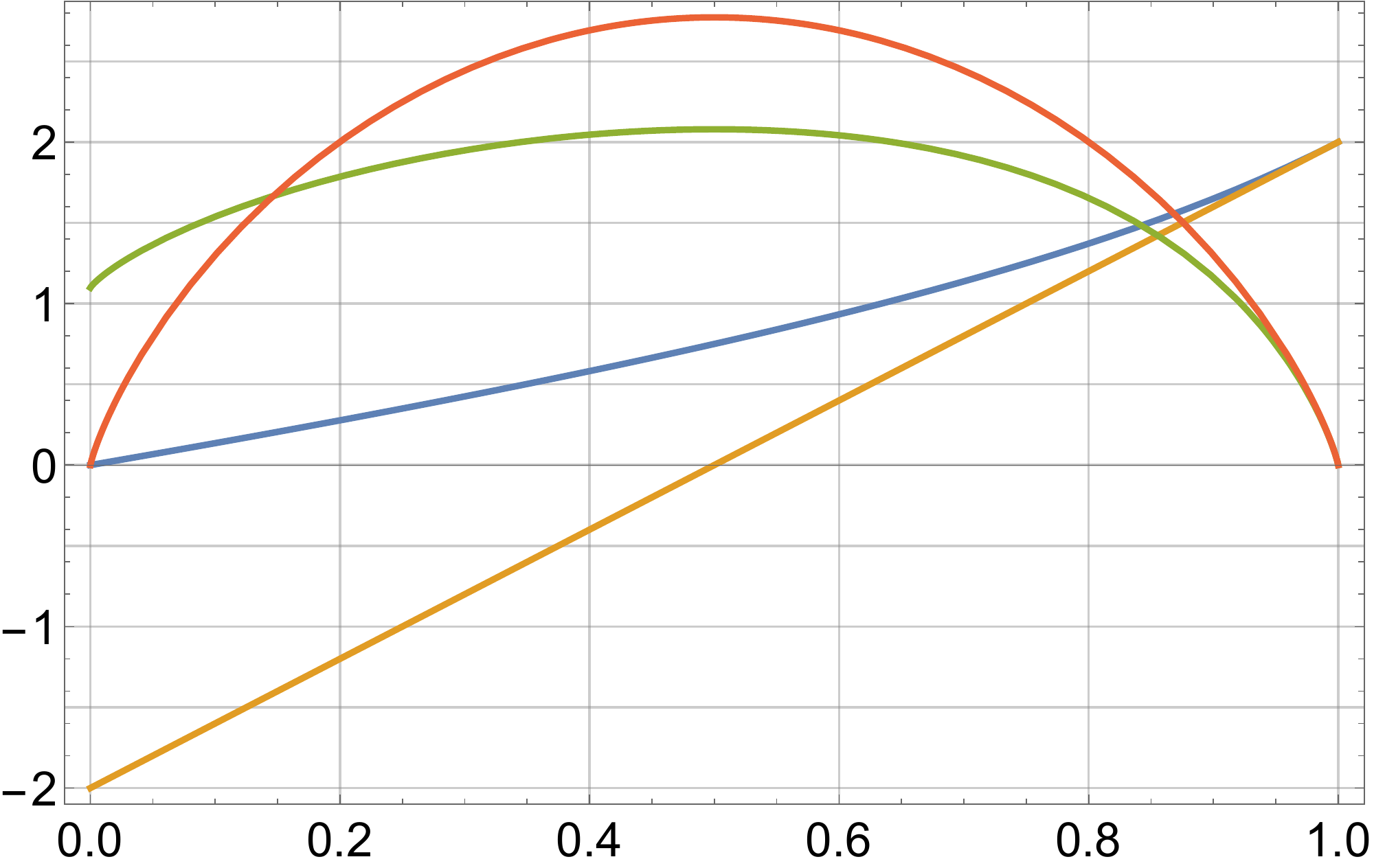}};
  \draw[red,thin] (-2.7,-2.1) -- (-2.7,2.5);
  \draw[black,very thick,|->] (-2.7,1.0) -- (-2.7,1.3);
  \draw[black,very thick,|->] (-2.7,-1.65) -- (-2.7,-0.01);
  \node[above=of img1, node distance=0cm, yshift=-2.7cm,xshift=-1.9cm] {{\color{black}$\Delta C^{(4-p)}$}};
  \node[above=of img1, node distance=0cm, yshift=-4.7cm,xshift=-2.cm] {{\color{black}$\Delta E^{(4-p)}$}};
  \node[above=of img1, node distance=0cm, yshift=-6.4cm,xshift=0cm] {$p$};
   \node[above=of img1, node distance=0cm, yshift=-2.cm,xshift=-2.cm] {{\color{black!40!red}$C_0^{(4)}$}};
   \node[above=of img1, node distance=0cm, yshift=-2.7cm,xshift=-.5cm] {{\color{black!60!green}$C_f^{(4-p)}$}};
   \node[above=of img1, node distance=0cm, yshift=-3.5cm,xshift=-1.7cm] {{\color{blue}$E_f^{(4-p)}/E$}};
    \node[above=of img1, node distance=0cm, yshift=-5cm,xshift=-0.5cm] {{\color{black!20!orange}$E_0^{(4)}/E$}};
\end{tikzpicture}
\caption{\label{fig-4-qub-pw}}
\end{subfigure}
\caption{(\subref{fig-coh}) The gain of coherence $\Delta C ^{(4)} >\Delta C ^{(3)}>\Delta C ^{(2)}$ increasing with the number $N$ of TLS constituting the battery. (\subref{fig-4-qub-pw}) The results for the pairwise approach, $N=4$ TLS, namely the plot of the normalized average energy $E_f^{(4-p)}/E$, $E_0^{(4)}/E$, the relative entropy of coherence $C_f^{(4-p)}$, $C_0^{(4)}$, Eq.~\eqref{eq-N-coh-pure-init}, for the final state $\op{P}_1^{\rm tot}\ket{\Psi_{i}^{(4)}}$, Eq.~\eqref{eq-pure-N-sequent}, and the pure initial state $\ket{\Psi^{(4)}_{i}}$, Eq.~\eqref{eq-N-Psii}. The values are plotted versus the single TLS excitation probability $p$,  Eq.~\eqref{eq-psi-pure}. The average energy gain, $0\leq E_f^{(4-p)}-E_0^{(4)}$ is achieved with maximum in the region of small $p$. The final entropy of coherence $C_f^{(4-p)}$ can be conditionally increased above the initial one $C_0^{(4)}$ in a region of small excitation probabilities $p$, whereas generally $\lim_{p\rightarrow 0}C_f^{(N-p)}=\ln(N/2+1)$, for even $N$. The red vertical line is a guide for the eye at $p=0.1$, and the black arrows show the energy and coherence increase.}
\label{fig-N-qub-pair}
\end{figure}

The examples of coherence, $C_f^{(N)}$ and $C_{0}^{(N)}$, are plotted in Fig.~\ref{fig-3-qub} and Fig.~\ref{fig-4-qub}. Their difference increases with $N$, contrary to the behavior of $\Delta E^{(N)}$, as well as the range of the excitation probabilities $p$ in which $\Delta C^{(N)}\equiv C_f^{(N)}-C_{0}^{(N)}\geq 0$. Notably, this last property is positive due to the fact that the higher $N$ we use, the less we are restricted by the choice of parameters for which we observe the positive effect of the charging protocol, i.e., the energy $E_f^{(N)}$ {\it and} coherence $C_f^{(N)}$ increase, over the respective initial values. Moreover, it follows from Eqs.~\eqref{eq-N-coh-pure-final}-\eqref{eq-N-coh-pure-init} that for fixed initial state excitation probability $p$, Eq.~\eqref{eq-psi-pure}, we obtain
\begin{equation}
\Delta C ^{(N)} >\cdots >\Delta C ^{(3)}>\Delta C ^{(2)},
\end{equation}
and the crossing points $\Delta C^{(N)}=0$ are shifting towards larger excitation probabilities $p$, see Fig.~\ref{fig-coh}. It is worth noting, that $\lim_{p\rightarrow 0}C_f^{(N)}(p)=\ln{N}$ is monotonically increasing with the number of TLS. However, taken relative to the maximum achievable coherence in the system of $N$ TLS, $C^{(N)}_{\rm max}=N\ln{2}$, the ratio $C_f^{(N)}(0)/C^{(N)}_{\rm max}$ approaches zero for increasing $N$. This result can be interpreted as the inability to fully exploit the values of coherence offered by increasing the dimension of our system with the present protocol in the region of $p\ll 1$. On the other hand, in the region $p\approx 1/2$ the post-measurement coherence value $C_f^{(N)}(1/2)$ approaches monotonically the value $C^{(N)}_{\rm max}$ with increasing $N$, c.f. Fig~\ref{fig-N-qub-global}. 

At this point we want to mention another possible charging protocol employing $N$ TLS. Its detailed analysis is beyond the scope of this paper.
It does not rely on {\it global} measurement on all TLS, but instead, on {\it sequential pairwise} application of the protocol introduced in the previous section, to the initial state \eqref{eq-N-Psii} in the sense $\op{P}_1^{\rm tot}\ket{\Psi_i^{(N)}}$, where
\begin{eqnarray}
\op{P}_1^{\rm tot}\equiv  \prod_{i=1}^{N-1}\op{P}_1^{(i-i+1)},\quad
\op{P}_1^{(i-j)}=\op{1}-\ket{g_ig_j}\bra{g_ig_j}.
\label{eq-pure-N-sequent}
\end{eqnarray}

Preliminary results suggest that such sequential pairwise protocol allows for energy increase, $\Delta E^{(N-p)}>0$, and coherence increase, $\Delta C^{(N-p)}>0$, in a qualitatively similar fashion as the global one, namely in the region of $p\ll 1$. On the quantitative side, the energy increase in the pairwise protocol is larger than in the global one, while for coherence it is the opposite case, c.f. Figs.~\ref{fig-4-qub-pw} and \ref{fig-4-qub} for numerical evidence. An example of this observation can be given for the zero excitation probability limit (even $N$) $\lim_{p\rightarrow 0}C_f^{(N-p)}=\ln(N/2+1)$, compared to its global counterpart $\lim_{p\rightarrow 0}C_f^{(N)}=\ln{N}$, Eq.~\eqref{eq-N-coh-pure-final}. The probability of success is in general smaller for the pairwise protocol compared to the global one, Eq.~\eqref{eq-pure-succ-N}. 

At this point, we would like to stress that such (simplified) protocol can be applied on all types of states (pure, dephased, after spontaneous emission, etc.) mentioned in the previous sections, due to its pairwise structure and it can be combined (in certain stages) with the repeat-until-success strategy introduced in the first section.

\section*{Conclusions and Outlook}
We have presented a protocol for the repeat-until-success charging of the quantum battery by means of quantum measurement application. The building blocks of our battery are factorized copies of identical two-level systems (TLS) with nonzero, although possibly very small, population of the excited state, and some small residual coherence that can be achieved by a suitable interaction with a sufficiently cold thermal bath\cite{giacomoPRL2018}. A pair of TLS subsequently undergoes a global, unity-resolving measurement with two possible outcomes, each represented by an operator diagonal in the TLS energy basis. 

If the protocol succeeds, we synthesize the initially independent pair of TLS into a system with higher coherent energy, thus creating and charging the battery simultaneously. The failure of the protocol results in reducing the initial energy and coherence to zero. These two possibilities represent the outcomes of the inherently conditional protocol, however with the possibility of increasing the success probability arbitrarily close to one (using repeat-until-success strategy), turning the protocol effectively into a deterministic one. Our results show that the partial purity and/or previous spontaneous emission of the energy of the initial state decreases the positive effect of the charging, but does not prevent the charging in principle. The variance of energy is decreased in this type of charging protocol.


The results of the projector-based charging protocol were generalized to the case of using $N$-TLS, either in a global or pairwise approach. The final value of the coherent energy can be increased even more, due to optimization, if the measurement consists of
POVM elements. Such protocol is superior in the value of coherence and energy with respect to the results of the projective measurement. The results should stimulate proof-of-principle experimental verification of such energy synthesizing using controlled quantum systems. To verify the observability of simultaneous energy and coherence increase, experimental tests can be implemented using photons\cite{BartuskovaPRA2006,MikovaPRA2013,CiampiniNPJQI2017,MancinoNPJQI2018}, employing such optical toolbox also with atoms\cite{MoehringNAT2007,SlodickaPRL2013,HofmannSCIENCE2012}, or solid-state systems\cite{ToganNAT2010,HensenNAT2015}. The implementation of measurement strategy for another experimental platforms is under development. This optical experimental tests would stimulate quantum thermodynamic analysis\cite{kammerlander,AlonsoPRL2016} of such synthesis, already used to analyze energy manipulations\cite{WalmsleyPRL2016,MancinoPRL2017,MancinoPRL2018}. 

We have verified that many paths are open to synthesize energy in coherent quantum battery for further optimization. This optimization is a demanding top-to-bottom task, therefore, it has to be solved numerically. 

\bibliography{sample}

\section*{Acknowledgements}
We gratefully acknowledge support of the project No. 19-19189S of the Grant Agency of the Czech Republic. R.F received national funding from the MEYS under grant agreement No. 731473 and from
the QUANTERA ERA-NET cofund in quantum technologies implemented within the European Union’s Horizon 2020 Programme (project TheBlinQC). R.F. and M.K. acknowledges CZ.02.1.01/0.0/0.0/16\_026/0008460 of MEYS CR. M.G. and R.F. acknowledges Palacky University IGA-PrF-2019-010.

\section*{Author contributions statement}
M.G. and M.K. did analytic calculations. R.F., M.G., and M.K. jointly analysed results. M.K. wrote manuscript with contributions from R.F. and M.G. R.F. developed the idea and supervised the project.

\section*{Competing interests}
The authors declare no competing interests.

\end{document}